\documentclass[12pt,indent]{article}
\usepackage{a4,latexsym,amsmath}
\usepackage[latin1]{inputenc}
\usepackage{float}
\usepackage{natbib}
\usepackage{graphics}
\usepackage[english]{babel}
\usepackage{tabularx}
\usepackage{lscape}
\usepackage{multirow}
\usepackage{rotating}
\usepackage{dsfont}
\usepackage{subfig}
\usepackage{bbm}
\usepackage{wrapfig}
\usepackage[table]{xcolor}

\usepackage{calc}
\usepackage{ifthen}

\newcommand{\E}{\operatorname{E}}      

\newcommand{\alphab}{\boldsymbol{\alpha}}

\newcommand{\betab}{{\boldsymbol{\beta}}}

\newcommand{\Sigmab}{\boldsymbol{\mathit\Sigma}}

\newcommand{\xb}{\boldsymbol{x}}

\newcommand{\zb}{\boldsymbol{z}}

\newcommand{\blanco}[1]{}

\def\d{\displaystyle}

\usepackage{natbib}

\usepackage{setspace}

\begin{document}
\bibliographystyle{chicago}
\sloppy

\makeatletter
\renewcommand{\section}{\@startsection{section}{1}{\z@}%
        {-3.5ex \@plus -1ex \@minus -.2ex}%
        {1.5ex \@plus.2ex}%
        {\reset@font\Large\sffamily}}
\renewcommand{\subsection}{\@startsection{subsection}{1}{\z@}%
        {-3.25ex \@plus -1ex \@minus -.2ex}%
        {1.1ex \@plus.2ex}%
        {\reset@font\large\sffamily\flushleft}}
\renewcommand{\subsubsection}{\@startsection{subsubsection}{1}{\z@}%
        {-3.25ex \@plus -1ex \@minus -.2ex}%
        {1.1ex \@plus.2ex}%
        {\reset@font\normalsize\sffamily\flushleft}}
\makeatother



\newsavebox{\tempbox}
\newlength{\linelength}
\setlength{\linelength}{\linewidth-10mm} \makeatletter
\renewcommand{\@makecaption}[2]
{
  \renewcommand{\baselinestretch}{1.1} \normalsize\small
  \vspace{5mm}
  \sbox{\tempbox}{#1: #2}
  \ifthenelse{\lengthtest{\wd\tempbox>\linelength}}
  {\noindent\hspace*{4mm}\parbox{\linewidth-10mm}{\sc#1: \sl#2\par}}
  {\begin{center}\sc#1: \sl#2\par\end{center}}
}



\def\R{\mathchoice{ \hbox{${\rm I}\!{\rm R}$} }
                   { \hbox{${\rm I}\!{\rm R}$} }
                   { \hbox{$ \scriptstyle  {\rm I}\!{\rm R}$} }
                   { \hbox{$ \scriptscriptstyle  {\rm I}\!{\rm R}$} }  }

\def\N{\mathchoice{ \hbox{${\rm I}\!{\rm N}$} }
                   { \hbox{${\rm I}\!{\rm N}$} }
                   { \hbox{$ \scriptstyle  {\rm I}\!{\rm N}$} }
                   { \hbox{$ \scriptscriptstyle  {\rm I}\!{\rm N}$} }  }

\def\d{\displaystyle}

\title{ Tree-Structured Modelling of Categorical Predictors in Regression}
\author{Gerhard Tutz \& Moritz Berger \\{\small Ludwig-Maximilians-Universit\"{a}t M\"{u}nchen}\\
{\small Akademiestra{\ss}e 1, 80799 M\"{u}nchen}}


\maketitle

\begin{abstract} 
\noindent
Generalized linear and additive models are very efficient regression tools but the selection of relevant terms becomes difficult  if higher order interactions are needed. In contrast, tree-based methods also known as recursive partitioning are explicitly designed to model a specific form of interaction but with their focus on interaction tend to neglect the main effects. The method proposed here focusses on the main effects of categorical predictors by using tree type methods to obtain clusters.
In particular when the predictor has many categories one wants to know which of the categories have to be distinguished with respect to their effect on the  response. The tree-structured approach allows to detect clusters of categories that share the same effect while letting other variables, in particular metric variables, have a linear or additive
effect on the response.  An algorithm for the fitting is proposed and various stopping criteria are evaluated. The preferred stopping criterion is based on $p$-values representing a conditional inference procedure. In addition, stability of clusters are investigated and the relevance of variables is investigated by bootstrap methods. Several applications show the usefulness of tree-structured clustering and a small simulation study demonstrates that the fitting procedure works well.

\end{abstract}


\section{Introduction}
Trees are a widely used tool in statistical data analysis. The most popular methods are CART, outlined in
\citet{BreiFrieOls:84},  and the C4.5 algorithm,  which was proposed by Quinlan (\citet{Quinlan:86},
\citet{Quinlan:93}).
An introduction into the basic concepts is found in \citet{HasTibFri:2009B},
an overview on recursive partitioning in the health sciences was given by \citet{ZhaSin:1999}  and an
introduction including random forests with applications in
psychology by \citet{Strobetal:2009}.

One big  advantage of trees is that they automatically find interactions. The concept of interactions is at the core of trees, which have its roots in  automatic interaction detection (AID), proposed by \citet{MorSon:63}. But the focus on interactions can also turn into a disadvantage because common trees do not allow for a linear or smooth component in the predictor. Below the root node most nodes represent interactions. Thus potentially linear or additive effects of covariates are hardly detected.

Moreover, in most regression problems one has a mixture of explanatory variables. Some are continuous, some are binary, others are categorical on a nominal scale or ordered categorical.
One application we will consider are the Munich rent standard data, which were also analysed in \cite{GerTut:2009a}. The data set consists of 2053 housholds with the  response variable being monthly rent per square meter in Euro. Available predictors  are the urban district (nominal factor), the year of construction, the number of rooms, the quality of residential area (ordinal factors), floor space (metric) and five additional binary variables. Conventional trees treat all these explanatory variables in a similar way. They split the predictor space by use of one variable into two regions. Within the regions the response is fitted as a constant. If in the first step a continuous explanatory variable is selected, for example floor space, in the next step typically interactions with floor space are fitted, more concise, interactions with the two selected regions of floor space.  In the next steps all fits refer to higher order interactions. Therefore, trees have a strong tendency  to fit interactions and neglect the main effects. Relevance of explanatory variables is found a posteriori by defining importance measures, which in random forests in some form reflect how often a variable has been selected, see, for example, \citet{ishwaran2007variable}, \citet{sandri2008bias} and \citet{Strobl-etal:2008}.
In contrast, a model that allows that a continuous variable has a smooth effect of unspecified functional form as in generalized additive models takes main effects much more serious. If, in addition, binary and categorical variables are included by use of a linear predictor   one obtains estimates of parameters that reflect the importance of the variables directly. The downside of generalized additive and parametric models is that higher order interactions are hard to model and that they can  contain a multitude of parameters.

The tree-structured approach proposed here uses trees in part of the variables but allows to include others as parametric or smooth components in the model. Our focus is on categorical predictors with many categories as, for example, the urban district in the rent data (25 districts). In particular categorical predictors are difficult to handle because for each category one parameter is needed. Thus simple parametric models tend to become unstable which calls for regularized estimates. Categorical predictors or factors come in two forms, unordered or ordered. In both forms one wants to know if the predictor has an impact, and, if it has, which categories have to be distinguished. The latter problem means that one wants to find  clusters of categories (or factor levels) which share the same expected response. In the nominal case all possible partitions of the set of categories are candidates, whereas in the ordered case clusters are formed by fusion of adjacent categories. The method proposed here uses trees to find the clusters of factor levels. Thus trees are used for the multi-categorical variables while the other variables are included in the classical form of linear or smooth effects.

Fusion of categories to obtain clusters of categories within a regression model has been mainly investigated by penalization methods, see \citet{BonRei:2009}, \citet{GerTutOrd:2009} and \citet{GerTut:2009a}. However, in contrast to the tree-structured approach, these  penalization techniques are restricted to a small number of categories. Penalization methods and  tree-type methods that are related or alternatives to the present approach  are considered in a separate section (Section \ref{sec:Related}). In Section 2 we introduce a tree-structured model for categorical predictors, in Section 3 the fitting procedure is presented. Section 4 deals with standard errors and the stability of clusters. A small simulation study is given in Section 6 and in  Section 7 we consider two further applications.

\section{Structured Predictors}

As in generalized linear models (GLMs) let the mean response $\mu=\E(y|\xb)$ be linked to the explanatory variables in the form
\[
\mu =h(\eta) \quad {\text or} \quad  g(\mu) = \eta,
\]
where $h(.)$ is the response function and $g(.)=h^{-1}(.)$ is the link function. As in GLMs we also assume that the distribution of $y|\xb$ follows a simple exponential family (\citet{McCNel:89}. While GLMs always assume that the predictor is linear we assume that the predictor is composed of two components, a tree component and a linear or additive component. With a linear component the predictor has the form
\[
\eta = tr(\zb)+\xb^T\betab,
\]
where $tr(\zb)$ is the tree component of the predictor and $\xb^T\betab$ is the familiar linear term.
Thus, one distinguishes between two groups of explanatory variables, namely $\zb$, which are determined by a tree, and $\xb$, which has a linear  effect on the response. In extended versions we consider the additive predictor
\[
\eta = tr(\zb)+\sum_{j=1}^q f_{(j)}(x_j),
\]
where the $f_{(1)}(.),\dots,f_{(q)}(.)$ are unspecified functions. Then one obtains a tree-structured model with additive components.

We will focus on the case where the $\zb$-variables are categorical. When a  tree is  built, successively
a node $A$, that is a subset of the predictor space, is split into subsets  with the split  determined by only one variable.
For a \textit{nominal} categorical variable $z \in \{1,\ldots,k\}$, the partition has the form
$A \cap S, \quad A \cap \bar S$,
where $S$ is a non-empty subset $S\subset\{1,\ldots,k\}$ and $\bar
S=\{1,\ldots,k\}\setminus S$ is the complement. Thus, after several splits the predictor $tr(z)$ represents a clustering of the categories $\{1,\ldots,k\}$, and the tree term can be represented by  \[
tr(z)= \alpha_1 I(z \in S_1)+\dots+\alpha_m I(z \in S_m),
\]
where $S_1,\dots,S_m$ is a partition of $\{1,\ldots,k\}$, and $I(.)$ denotes the indicator function with $I(a)=1$ if $a$ is true, $I(a)=0$ otherwise.

For an \textit{ordinal} categorical variable $z \in \{1,\ldots,k\}$ the partition into two subsets has the form
$A \cap \{ z \leq c \}, \quad A \cap \{ z > c \}$,
based on the threshold $c$ on variable $z$. Thus  during the building of a tree clusters of adjacent categories are formed. The tree term has the same form as before but with the subsets that represent the clusters having the form $S_r=\{a_{r-1}, \dots,a_{r}\}$, $a_{r-1} <a_{r}$.

In the case of more than one categorical predictor the tree-structured clustering proposed here  forms clusters only for one variable. Then, with $p$ predictors in $\zb$ the tree component has the form
\[
tr(\zb)= tr(z_1)+\dots+tr(z_p),
\]
where $tr(z_r)$ is the tree for the $r$th variable, that means it represents clusters of the $r$th variable with the cluster form determined by the scale level of the corresponding variable.
A traditional tree hardly   finds clusters for single components. It typically produces clusters that combine several variables, in particular, mixing nominal and ordinal predictors.

Clustering by trees  is a forward selection strategy. But one should be aware that the all subsets strategy fails even in cases of a moderate number of categories. Already in the  case of only one predictor one has to consider all subsets $S_1,\dots,S_m$ and fit the corresponding model with predictor $\eta=\alpha_1 I(z \in S_1)+\dots+\alpha_m I(z \in S_m)+\xb^T\betab$. This is computational feasible only for a very small number of categories. For  more than one variable one has to consider all possible combinations, which is bound to fail.

\section{Tree-Structured Clustering}

For simplicity we start  only one categorical predictor and consider the general case in later sections.
\subsection{Trees with Clusters in a Single  Predictor}

\subsubsection*{Ordered Categorical or Metric Predictor}
Let us start with one ordinal or metric variable $z$. Then one split in a tree that includes a linear predictor is found by fitting a model with predictor
\[
\eta = \alpha_l I(z \le c)+ \alpha_r I(z > c)+ \xb^T\betab,
\]
where $I(.)$ again denotes the indicator function. By use of the split-point $c$ the model splits the predictor space into two regions, $z \le c$ and $z > c$. In the left node, for all $z \le c$, one specifies the response level $\alpha_l$, in the right node, for all $z > c$, one specifies the level $\alpha_r$. It should be emphasized that in $\xb$  no  intercept is included. An equivalent representation of the predictor is
\[
\eta = \beta_0+ \alpha I(z > c)+ \xb^T\betab.
\]
with the transformation of parameters given by $\beta_0=\alpha_l$ and $\alpha=\alpha_r-\alpha_l$. The latter form of the predictor is more convenient since it contains an intercept as common regression models do and only one step function has to be specified.

When growing trees one has to specify the possible split-points. Let in the following $C=\{c_1,\dots, c_m\}$ denote the possible splits. For a metric predictor, in principle all possible thresholds $c$ can be used, but it suffices to use as candidates all the distinct observations available for  the predictor.  Therefore, $C$ contains the distinct values of the observed predictor. For ordinal predictors $z \in \{1,\ldots,k\}$ the set $C$ is simply $\{1,\dots, k\}$ and $m=k$.

The basic algorithm that we are using for an ordinal or metric variable is the following.

\vspace{0.5 cm}
\hrule
\begin{center}{\bf Tree-Structured Clustering - Single Ordered Predictor}\end{center}

\begin{description}
\item{\it Step 1 (Initialization)}

\begin{itemize}
\item[(a)] Estimation: Fit the candidate GLMs with predictors
\[
\eta = \beta_0+ \alpha_j I(z > c_j)+ \xb^T\betab, \quad j=1,\dots, m
\]
\item[(b)] Selection

Select the model that has the best fit. Let $c_{j_1}^*$ denote the best split.
\end{itemize}

\item{\it Step 2 (Iteration)}

For $l=1,2,\dots$,

\begin{itemize}
\item[(a)] Estimation: Fit the candidate models with predictors
\[
\eta = \beta_0+ \sum_{s=1}^l\alpha_{j_s} I(z > c_{j_s}^*) + \alpha_{j} I(z > c_{j}) + \xb^T\betab,
\]
for all values $c_{j} \in C \setminus \{c_{j_1}^*,\dots,c_{j_l}^*\}$

\item[(b)] Selection

Select the model that has the best fit yielding the cut point $c_{j_{l+1}}^*$.
\end{itemize}
\end{description}
\hrule
\vspace{0.5 cm}

The algorithm uses two steps, fitting of candidate models and selection of the best model. In GLM-type models it is quite natural to measure the fit by the deviance. Thus, one selects the model that has the smallest deviance. The criterion is equivalent to minimizing the entropy, which has been used as a splitting criterion already in the early days of tree construction (\citet{BreiFrieOls:84}).

The algorithm  yields a sequence of fitted split-points $c_{j_1}^*, c_{j_2}^*, \dots$ from $C$ and the corresponding parameter estimates  $\hat\alpha_{j_1}, \hat\alpha_{j_2}, \dots$ from the last fitting step. Typically the selection of split-points is stopped before all possible splits are included (for stopping criteria see below) and one obtains the subset of selected splits $C^*=\{c_{j_1}^*\dots,c_{j_L}^*\}$, where $L$ denotes the number of selected split-points.  Since the fitted functions are step functions one obtains  a partitioning into clusters of adjacent categories. For ordered categories the thresholds are given by $C=\{1,\dots, k\}$ and one obtains the clustering after ordering  the selected thresholds such that   $c_{(j_1)} \le c_{(j_s)} \le \dots$ by $\{1,\dots, c_{(j_1)}\}, \{c_{(j_1)}+1,\dots, c_{(j_2)}\} \dots$.  If in the initialization step the maximal value from the set of considered split-points, $C$, is selected, the algorithm stops immediately because in the iteration steps always the same model would be found. Then, $\hat\alpha_1=0$ and no split-point is selected; the variable is not included.

Although the method generates  trees   the methodology differs from the fitting of common trees if a parametric term is present.
In common trees without a parametric term partitioning of the predictor space is equivalent to splitting the set of observed data accordingly. In the next split only the data
from the corresponding subspace are used. For example, when a split yields the partition $\{z\le c \}$,  $\{z > c \}$, in the next split only the data from $\{z\le c \}$ (or $\{z > c \}$) are used to obtain the next split. This is different for the tree-structured model. In all of the fitting steps all data are used. This ensures that one obtains valid estimates of the parametric component together with the splitting rule. Of course, if no parametric component is present, the algorithm is the same as for common trees.

The method  explicitly does not use off-sets. When fitting within the iteration steps the previously fitted models serve only to specify the split-points that are included in the current fit.
But no estimates from the previous steps are kept. This is in contrast to \citet{yu2010partially}, where off-sets are used.

\subsubsection*{Stopping Criterion}

When building a tree it is advisable to stop after an appropriately chosen number of steps. There are several strategies to select the number of splits. One strategy that has been used since the introduction of trees is to grow large trees and prune them afterwards, see
\citet{BreiFrieOls:84}  or \citet{Ripley:96}, Chapter 7. Alternative strategies based on conditional inference procedures were given by \citet{Hotetal:2006}.

We use as one strategy \textit{k-fold cross-validation}. That means the data set is split into k subsets. The tree is grown on $k-1$ of these subsets, which is considered the  learning sample, and then the tree is evaluated on the left-out sub sample. Since we are working within the GLM framework a natural candidate for the evaluation criterion is the predictive deviance. The number of splits that showed the best performance in terms of the predictive deviance is chosen in the final tree fitted for the whole data set.

An alternative is to use a \textit{stopping criterion based on p-values}, a procedure that is strongly related to the conditional inference procedure proposed by  \citet{Hotetal:2006}. In each step of the fitting procedure one obtains a $p$-value  for the parameter that determines the splitting. In our notation, in the $l$th split one tests the null hypotheses $H_0: \alpha_l =0$ yielding the $p$-value $p_l$ for the selected split. Typically the sequence of $p$-values $p_1,p_2,\dots$ is increasing. A simple criterion is to stop if the $p$-values are larger than a pre-specified threshold $\alpha$. However, one should adapt for multiple testing errors because
in each split several hypotheses are tested. A simple strategy is to use the Bonferroni procedure and stop if $p_l > \alpha/(m-(l-1)) $ because in the $l$th split $m-(l-1)$ number of parameters are tested.  Then, in each step the overall error rate is under control. As test statistic one can use the Wald statistic or the likelihood ratio statistic. Although the Wald statistic is easier to compute, we prefer the likelihood ratio statistic because it corresponds to the selection criterion, which selects the model with minimal deviance.

\subsubsection*{Nominal Predictor}

For a nominal predictor $z \in \{1,\ldots,k\}$ splitting is much harder because one has to consider all possible partitions that contain two subsets. That means one has $2^{k-1}-1$ candidates for splitting. For large $k$ the number of candidates is excessive. But it has been shown that for regular trees it is not necessary to consider all possible partitions. One simply orders the predictor categories by increasing mean of the outcome and then splits the predictor as if it were an ordered predictor. It has been shown that this gives the optimal split in terms of various split measures, see \citet{BreiFrieOls:84} and \citet{Ripley:96} for binary outcomes and \citet{fisher1958grouping} for quantitative outcomes and the remarks of \citet{HasTibFri:2001}.

\subsection{Trees with Clusters in More than One Predictor}

If several predictors are included in the tree component the algorithm also selects among the available variables.
Let $C_i$ denote the possible splits in variable $z_i$ and $m_i$ denote the number of values in $C_i$.
The basic form of the algorithm is the following.

\vspace{0.5 cm}
\hrule
\begin{center}{\bf Tree-Structured Clustering - Several Ordered Predictors}\end{center}

\begin{description}
\item{\it Step 1 (Initialization)}

\begin{itemize}
\item[(a)] Estimation: Fit the candidate GLMs with predictors
\[
\eta = \beta_0+ \alpha_{ij} I(z_i > c_{ij})+ \xb^T\betab, \quad i=1,\dots,p, j=1,\dots, m_i
\]
\item[(b)] Selection

Select the model that has the best fit. Let $c_{i_1, j_1}^*$ denote the best split, which is found for variable $z_{i_1}$.
That means that $c_{i_1, j_1}^*$ is from the set of possible splits for  $z_{i_1}$.

\end{itemize}

\item{\it Step 2 (Iteration)}

For $l=1,2,\dots$,

\begin{itemize}
\item[(a)] Estimation: Fit the candidate models with predictors
\[
\eta = \beta_0+ \sum_{s=1}^l\alpha_{i_s, j_s } I(z_{i_s} > c_{i_s, j_s}^*) + \alpha_{ij} I(z_i > c_{ij}) + \xb^T\betab,
\]
for all $i$ and all values $c_{ij} \in C_i$  that have not been selected in previous steps.

\item[(b)] Selection

Select the model that has the best fit yielding the new cut point $c_{i_{l+1},j_{l+1}}^*$ that is found for variable $z_{i_{l+1}}$.
\end{itemize}
\end{description}
\hrule
\vspace{0.5 cm}

In the sequence of selected cut-points $c_{i_1, j_1}^*, c_{i_2, j_2}^*,\dots$ and corresponding estimates $\hat\alpha_{i_1, j_1}, \hat\alpha_{i_2, j_2},\dots$ the first index refers to the
variable and the second to the split for this variable. The selected splits for the $i$th variable can be collected in $C_i^*$, which comprises all splits $c_{i_l, j_l}^*$for which $i_l=i$ holds.

\begin{figure}[!ht]
\centering
\includegraphics[width=0.8\textwidth]{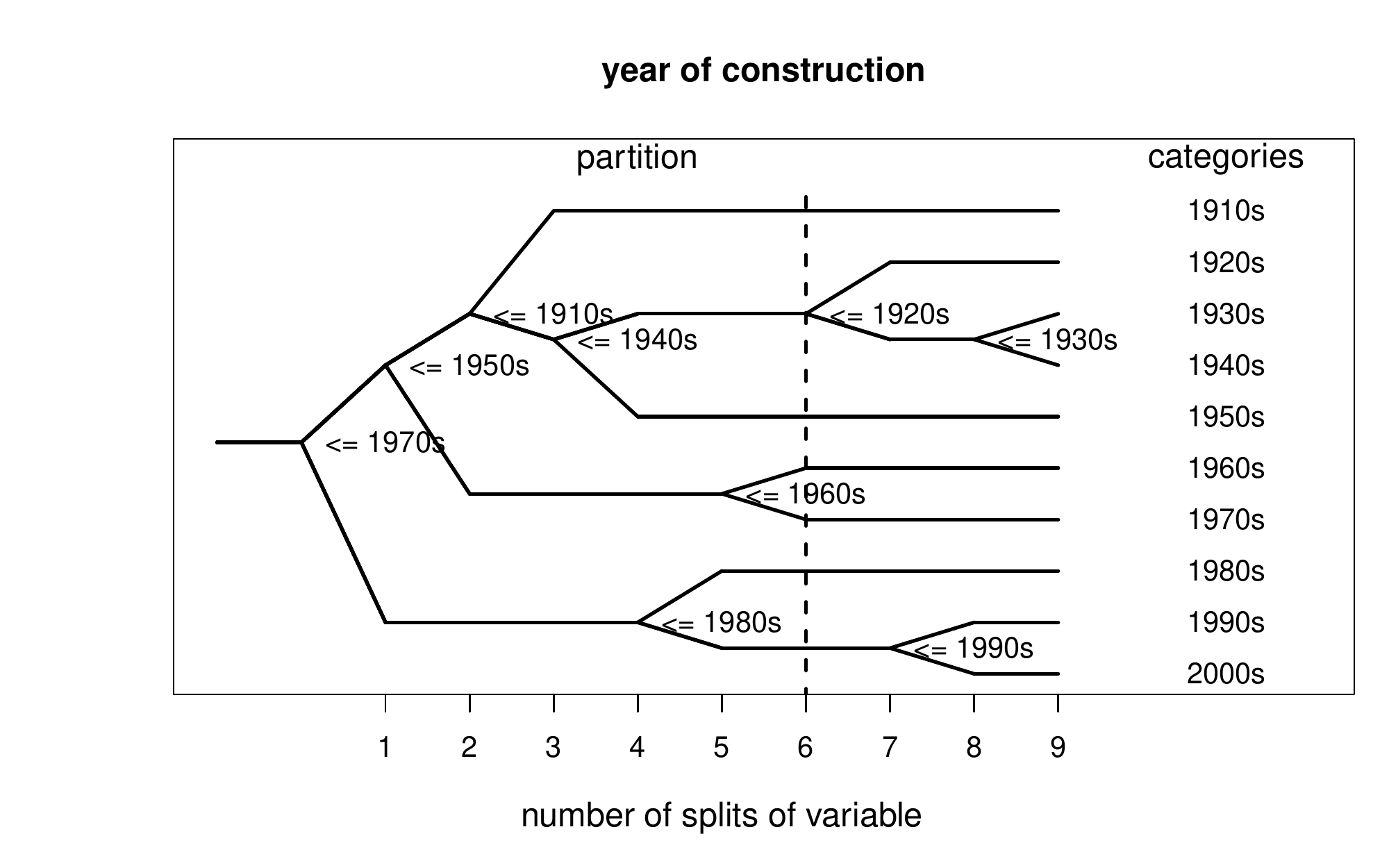}
\includegraphics[width=0.8\textwidth]{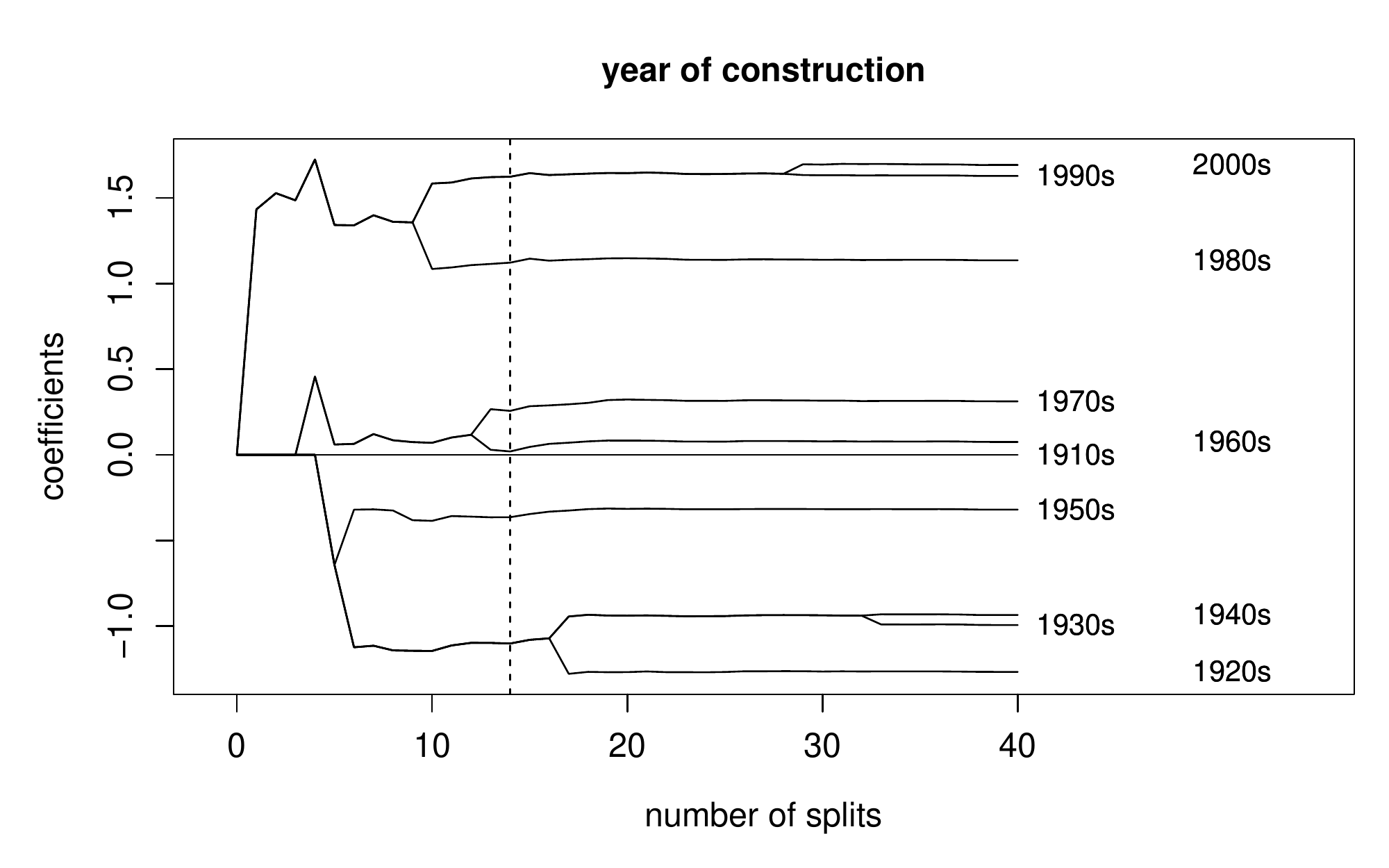}
\caption{Results for the ordinal predictor year of construction for the analysis of the Munich rent standard data. Upper panel: resulting tree for  year of construction, lower panel: paths of coefficients against all splits.}
\label{fig:bj_miete}
\end{figure}

\begin{table}[!ht]
\centering
\begin{footnotesize}
\begin{tabularx}{\textwidth}{Xcrr}
\hline
\bf{Predictor}&\bf{Cluster}&\bf{Coefficient}&\bf{Stabilty}\\
\hline
&&&\\
Urban district&7,11,14,16,22,23,24&-1.525&0.431\\
&6,8,10,15,17,19,20,21,25&-1.005&0.421\\
&9,13&-0.647&0.506\\
&2,4,5,12,18&-0.368&0.511\\
&1,3&0.000&0.552\\
&&&\\
Year of construction&1910&0.000&1.000\\
&1920s,1930s,1940s&-1.098&0.730\\
&1950s&-0.365&1.000\\
&1960s&0.030&1.000\\
&1970s&0.267&1.000\\
&1980s&1.115&1.000\\
&1990s,2000s&1.622&0.927\\
&&\\
Number of rooms&1,2,3&0.000&0.642\\
&4,5,6&-0.327&0.865\\
&&\\
Quality of residential area&fair&0.000&1.000\\
&good&0.356&1.000\\
&excellent&1.436&1.000\\
\hline
\end{tabularx}
\vspace{0.5cm}

\begin{tabularx}{\textwidth}{Xrr}
\hline
\bf{Predictor}&\bf{Coefficient}&\bf{95$\%$ confidence interval}\\
\hline
&&\\
Hot water supply (no)&-1.987&[-2.513,-1.372]\\
Central heating (no)&-1.355&[-1.820,-0.947]\\
Tiled bathroom (no)&-0.543&[-0.786,-0.318]\\
Supplementary equiment in bathroom (yes)&0.511&[0.199,0.807]\\
Well equipped kitchen (yes)&1.198&[0.839,1.579]\\
\hline
\end{tabularx}
\end{footnotesize}
\caption{Estimated coefficients, stability measures of the tree component and $95\%$ confidence intervals of the linear term for the analysis of the Munich rent standard data with a optimal number of 13 splits in the tree component.}
\label{tab:estimate_miete}
\end{table}

\begin{figure}[!ht]
\centering
\includegraphics[width=0.8\textwidth]{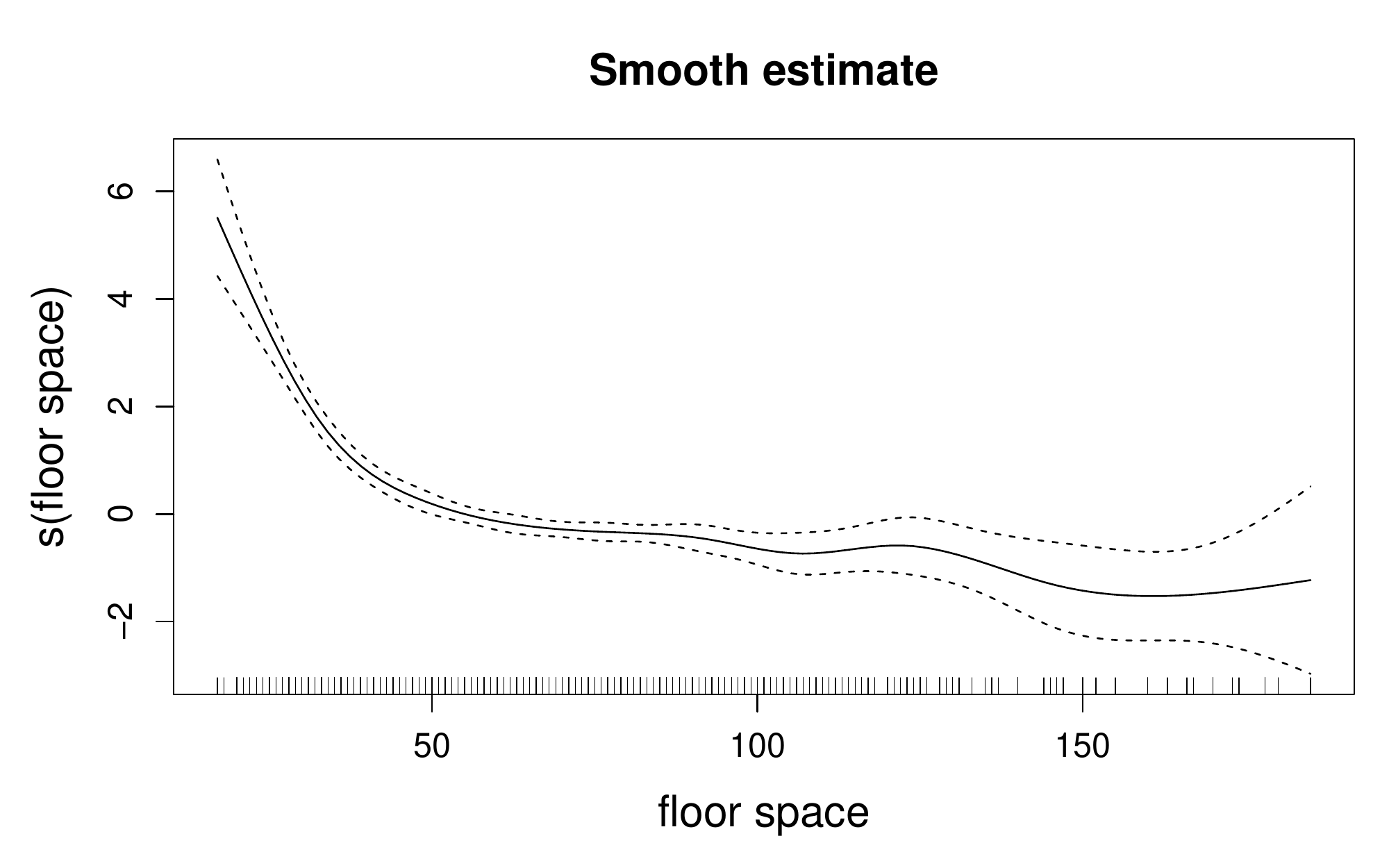}
\caption{Resulting function of the smooth estimation of predictor floor space of the Munich rent  data in the additive part of the  tree.}
\label{fig:miete_wfl_smooth}
\end{figure}

\subsection{Trees for Rent Data}
In the Munich rent data one has one nominal predictor (urban district), three ordinal predictors (year of construction in decades, number of rooms, quality of residential area), one metric variable (floor space) and five  binary variables. In the additive part  we model the effect of the metric predictor by cubic regression splines and include the binary variables in a linear form. The fusion of categories obtained by the tree is illustrated for the predictor year of construction. Figure \ref{fig:bj_miete} shows the resulting tree and the coefficient paths over the  splits   for the predictor decade of construction. The upper panel shows the successive splits  against the number of splits in this predictor. The lower panel shows the coefficients  plotted against  the splits in all of the  predictors.  It is seen, in particular from the first steps, that estimates can change when other variables are included. But after about 14 splits the estimates are very stable. Since the maximal number of splits is 40 the estimates after 40 splits represent the fit of a partial additive model. When $p$-values with significance level 0.05 are used as splitting criterion one obtains seven clusters  marked by the dashed lines in both panels. The rent per square meter seems to be the same, for example,  for houses built between the 1920s and 1940s and  for houses built in the 1990s and 2000s. The  gap between the high rent cluster and the middle clusters is larger than the gap between the middle clusters and the low rent clusters. The estimated values are given in Table \ref{tab:estimate_miete}, which also shows the clusters for the other variables. The sizes of clusters found by the algorithm vary in a wide range. The smallest clusters for the variable urban district consist of only one category, the biggest cluster contains 9 categories. For the variable year of construction the biggest cluster contains three categories. It should be noted that  no predictor has been completely excluded. Table \ref{tab:estimate_miete} also contains stability measures that are explained later. Since it is not to be expected that the rent per square meter depends linearly on the floor space it is fitted as a  smooth function.

For the estimation we use penalized cubic regression splines, penalized by the  integrated squared second derivative penalty (\citet{EilMar:96}). We chose a modest number of ten basis functions; for computation we used the R-package mgcv (\citet{wood2011mgcv}). When fitting a smooth function one has to specify a smoothing parameter, which in our procedure  is selected new in each iteration step.  The resulting function, pictured in Figure \ref{fig:miete_wfl_smooth}, is monotonically decreasing, which  means that the net rent per square meter decreases with growing floor space. The function decreases strongly until a floor space of about 50 and is rather flat for a greater floor space, but it is definitely not linear.

\begin{figure}[h!]
\centering
\includegraphics[width=0.8\textwidth]{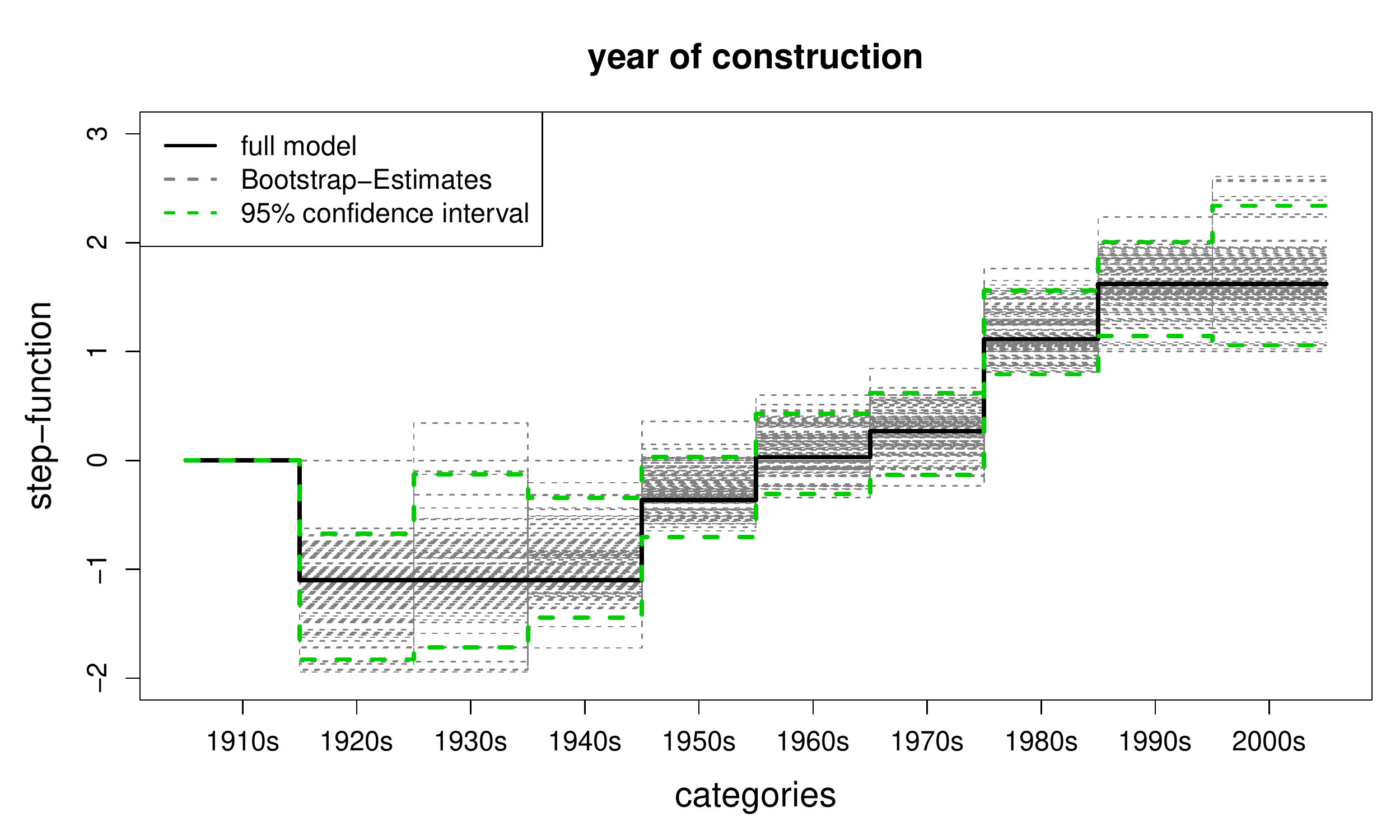}
\includegraphics[width=0.8\textwidth]{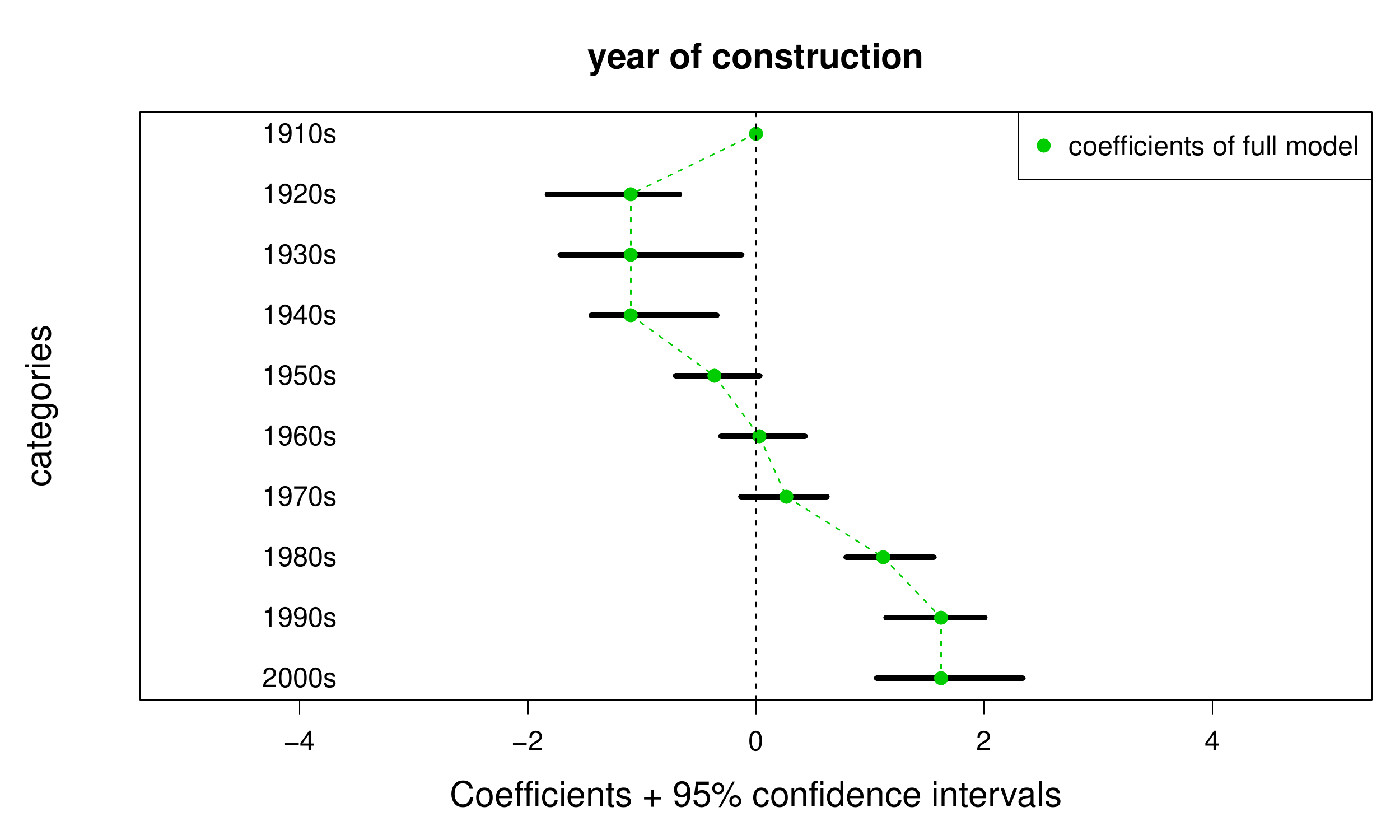}
\caption{Estimated step functions and resulting $95\%$ confidence intervals for the ordinal predictor year of construction for the analysis of the Munich rent standard data based on 1000 bootstrap samples.}
\label{fig:bj_miete_KI}
\end{figure}

\begin{figure}[h!]
\centering
\includegraphics[width=0.8\textwidth]{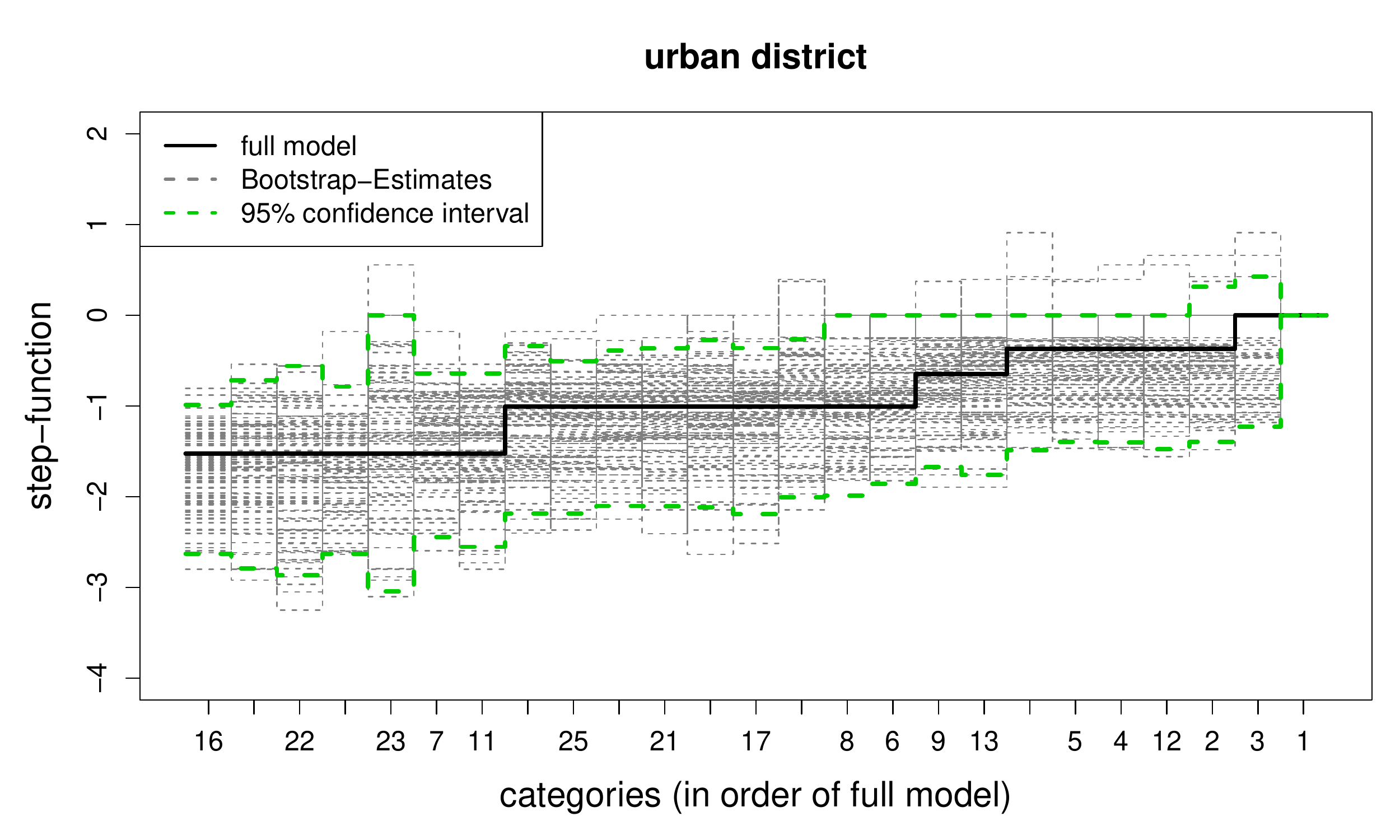}
\includegraphics[width=0.8\textwidth]{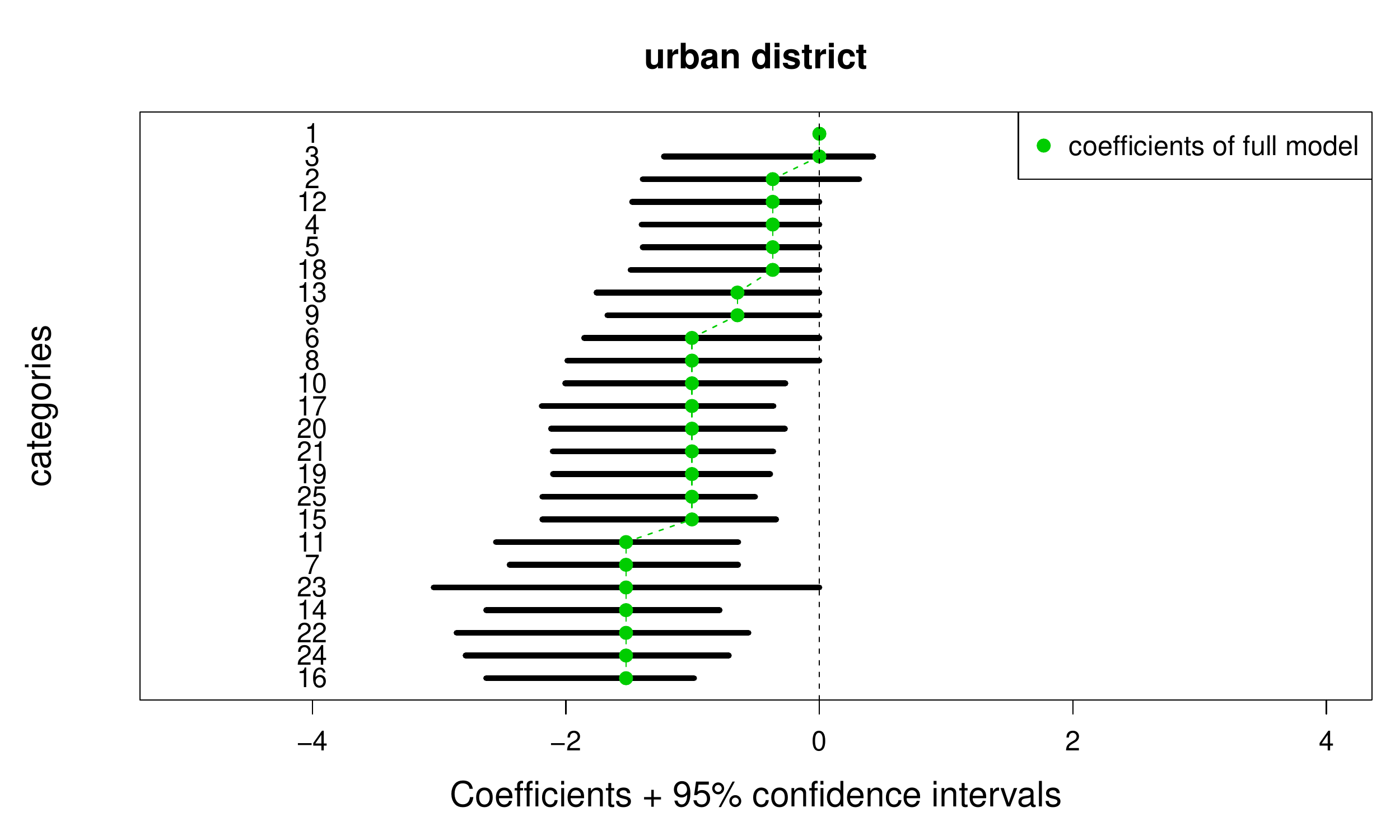}
\caption{Estimated step functions and resulting $95\%$ confidence intervals for the nominal predictor urban district for the analysis of the Munich rent standard data based on 1000 bootstrap samples.}
\label{fig:bez_miete_KI}
\end{figure}

\section{Standard Errors and Stability of Clusters}

The tree-structured  model  is an  extension of  GLMs and GAMs. While in standard GLMs approximate standard errors for the parameters are obtained from asymptotic theory,
for semiparametric models as considered here an alternative way to obtain standard errors has to be used. One way is to use bootstrap procedures as described in \citet{efron1994introduction}. By repeated fitting on sub samples that have been obtained by drawing with replacement one can compute approximate standard errors.
But when computing standard errors one has to distinguish between the two parts of the model, the parametric and the tree part. For the parametric part, which means for the parameter $\betab$, standard procedures to compute the standard deviations and confidence intervals over the bootstrap samples can be used. For the rent data the resulting confidence intervals are given in Table \ref{tab:estimate_miete}.  For  categorical predictors we consider the estimated step functions, which are determined by sums of
the parameter estimates $\hat\alpha_{ij}$. Bootstrap intervals can be given for all estimated sums $\tilde\alpha_{is}=\sum_{j=1}^s\hat\alpha_{ij}$. Typically some of the parameter estimates $\hat\alpha_{ij}$  are zero, but this has not to be the case  in the bootstrap samples. Consequently one obtains  confidence intervals that do not necessarily have equal length within clusters.
The somewhat harder problem is the case of nominal predictors. Since in bootstrap  samples the ordering of the predictor categories will differ one has to carefully rearrange the  parameter estimates  to obtain the confidence intervals for the estimates $\tilde\alpha_{is}$ in the original sample.

For illustration we show the bootstrap results for the variables year of construction (\ref{fig:bj_miete_KI}) and urban district (\ref{fig:bez_miete_KI}).
The upper panels of Figure \ref{fig:bj_miete_KI} and Figure \ref{fig:bez_miete_KI} show only the first 100 bootstrap based  function  estimates. The lower panels show the $95\%$ confidence intervals for the single effects for 1000 bootstrap samples.
It is seen that for year of construction the first big  cluster, which contains the  decades 1920-1940, has varying lengths of confidence intervals, but all of them do not contain zero.
Thus they should be distinguished from the reference category, which is the first decade, and has fixed value zero. For the nominal predictor urban district  the reference category is district one, which denotes the inner city of Munich (around Marienplatz). As was to be expected for a nominal variable with many categories confidence intervals are larger than for the variable years of construction. But it is seen that several big clusters are definitely less expensive than the district inner city.

Bootstrapping yields confidence intervals for the step functions but do not contain information about the reliability of cluster identification. Therefore it seems warranted to supplement the confidence intervals by diagnostic tools that reflect the stability of clusters. One is a distance matrix obtained from the bootstrap samples. Let $B$ denote the number of bootstrap samples and $n_{ij}$ denote the number of samples for which category $i$ and $j$ were in the same cluster. Then a simple  similarity measure for categories is $s_{ij}= n_{ij}/B$. If $s_{ij}=1$ category $i$ and $j$ were in the same cluster in all of the bootstrap samples. The stability of a cluster is obtained by averaging over all the distances of pairs of categories within  the cluster. Of course, if a cluster contains only one category the similarity measure has the value 1. It is seen from Table \ref{tab:estimate_miete} that stability can strongly vary across clusters. For the nominal variable urban district the clusters show similarity in the range $(0.43,0.53)$ whereas for the ordinal variable year of construction one obtains also very large values as 0.73 and 0.927. The latter value refers to the cluster of decades 1990 and 2000 and means that it was in the same cluster in 92.7\% of the bootstrap samples.

\section{Related Approaches}\label{sec:Related}

In the following the relation of the proposed method to other methods that have been proposed is shortly sketched. Our method  aims at the identification of clusters in categorical predictors in the presence of other, in particular, also continuous variables. Therefore discussion refers to this objective.

The strongest relation is to classical recursive partitioning or simple trees as CARTS. The main differences have already been outlined. The method proposed here allows to include a parametric or smooth component that accounts for the main effect in a model. Thus, the method allows to identify clusters of categories within one predictor that have the same effect on the response. If one fits a classical tree that includes all the variables no clustering is obtained because the tree fits interactions between all the variables.

As a forward strategy one might suspect a strong relation to boosting concepts.
Boosting methods were originally developed in the machine
learning community as a means to improve classification (e.g.,
\citet{Shap:90}). Later it was shown that
it can be seen as the fitting of an additive structure by
minimizing specific loss functions (see
\citet{Friedman:2001},
\citet{FriHasTib:2000},
\citet{BueYu:2003}). Minimization is obtained iteratively by utilizing a steepest gradient descent approach. In a forward searching procedure  components that are potentially relevant are included in the predictor. The potentially relevant components are fitted by so-called base learners. A simple example is the fitting of a linear model where a base learner refers to the fitting of one component of the linear predictor, $x_i\beta_i$. By including one of the components at a time and selection of the component that maximally improves the fit one obtains the final model. The method proposed here seems to be very similar. The base learner that is used is one split in a variable, which has the form   $\alpha_{ij} I(z_i > c_{ij})$. Selection of the most relevant term is also based on goodness-of-fit. However, there is one crucial difference between the  tree-structured clustering and boosting, namely that boosting uses weak learners. A weak learner is somehow vaguely defined as a refit that only slightly improves the overall fit, but properties of the procedure definitely depend on the weakness of the learner (\citet{BueYu:2003}). In our procedure a weak learner would be the inclusion of the best split $\alpha_{ij} I(z_i > c_{ij})$, but with a new parameter value $\alpha_{ij}$ that is only slightly larger than the parameter used in the previous step. Of course, one could fit categorical predictors by weak learners, or equivalently by boosting,  but the effect would be a smooth fit over categories, because in each boosting step the parameters are updated only weakly but most of them are selected during the iterations. Therefore, the procedure fails to obtain the intended clustering of categories.

Approaches that are able to detect  clusters in categories can be constructed by the definition of appropriate penalty terms and maximization of the corresponding
penalized log-likelihood. Let us for simplicity consider the case of one categorical predictor and several continuous predictors. Then the corresponding linear predictor of a GLM is given by
\[
\eta = \alpha_0+\alpha_1 \tilde z_1+\dots +\alpha_{k-1} \tilde z_{k-1}+\xb^T\betab,
\]
where $\tilde z_j$ are the dummy variables for the categorical predictor $z\in \{1,\dots,k\}$. Let the penalized log-likelihood be given by $l_p(\alphab,\betab)= l(\alphab,\betab) - J(\alphab,\betab)$,
where $l(\alphab,\betab)$ denotes the log-likelihood of the GLM and $J(\alphab,\betab)$ is a penalty term. For a categorical predictor a penalty that enforces clustering of categories of $z$
is given by
\[
J(\alphab,\betab)= \lambda\sum_{r<s} |\beta_r-\beta_s|,
\]
For $\lambda=0$ one obtains the ML estimate, if $\lambda \rightarrow \infty$ all categories of $z$ are fused to one cluster. The method has been proposed by \citet{BonRei:2009} for ANOVA-type models and was adapted to variable selection by  \citet{GerTut:2009a}, \citet{TutGer:2014}. With appropriate tools to select the tuning parameter $\lambda$ the method shows good performance in detecting clusters, in particular in extended versions that include weights in the penalty term. The main problem with this approach is that it becomes computationally infeasible if the number of categories gets large. This is due to the definition of the penalty term, which includes all pairwise differences. If the number of categories is 40, the penalty already contains 780 differences.

A further approach  that is related to the tree-structured model is  model-based partitioning proposed by
\citet{zeileis2008model}. The basic concept is to fit a parametric model in every leaf of a tree,  for example,  a linear regression model.
By fitting a model to subsets that are defined in the usual way by splitting variables one obtains a partitioned
or segmented parametric model.  Within this  framework it is possible to detect areas where model fits differ because the linear models fitted to leafs differ in their parameters.
It is a flexible modeling tool in which all kinds of parametric models can be used. However, as in common trees the focus is not on main effects but on  interaction although  in the wider sense that models differ in different leafs. In particular for categorical predictors, which are considered here, one obtains different structures when using model-based partitioning in the sense of \citet{zeileis2008model} or  structured regression as proposed here. In model-based partitioning  splits  in a categorical predictor are enforced if the parameters of the fitted model differ in the resulting clusters of categories. After several splits one obtains quite different models that hold within clusters of categories. In our structured regression clusters of categories are built  by assuming that the effect on the response is the same within clusters and that the main effects are constant. Thus the focus is on similarity of categories not on dissimilarity of categories with respect to the models that hold within clusters of categories.

Finally, several modelling strategies were proposed that also use a combination of a parametric term and a tree component. One is the partially linear tree-based regression model developed by \citet{chen2007partially}. The focus of the paper is on genetic risk factors. The main difference to the procedure proposed here is the restriction to a linear term and an alternative algorithm that uses off-sets in the iterative algorithm instead of updating the linear component. The approach has been extended to account for multivariate outcomes by \citet{yu2010partially}. An alternative model   is the regression trunk model proposed in \citet{dusseldorp2004regression} and \citet{dusseldorp2010combining}. The model is designed for metric response only. In contrast to our approach it  uses the same variables in the tree component and the linear term, which yields hard to interpret effects. Moreover, they use the more conventional fitting strategy that first grows a large tree and then prunes it. Therefore, the relevance of predictors in terms of significance should be hard to obtain. A combination of linear fits and tree-structured component with the focus on diagnostic for linear models was considered by \citet{su2009tree}.

\section{Simulations }

One of the most important questions when building a tree is the choice of a optimal stopping criterion. In our applications we used a stopping criterion based on $p$-values of the likelihood-ratio statistic. To investigate the performance of the tree-structured clustering, especially the ability to form correct clusters, and to compare  stopping criteria, we  give the results of a small simulation study.

\begin{figure}[!ht]
\centering
\begin{tabular}{cc}
\includegraphics[width=0.5\textwidth]{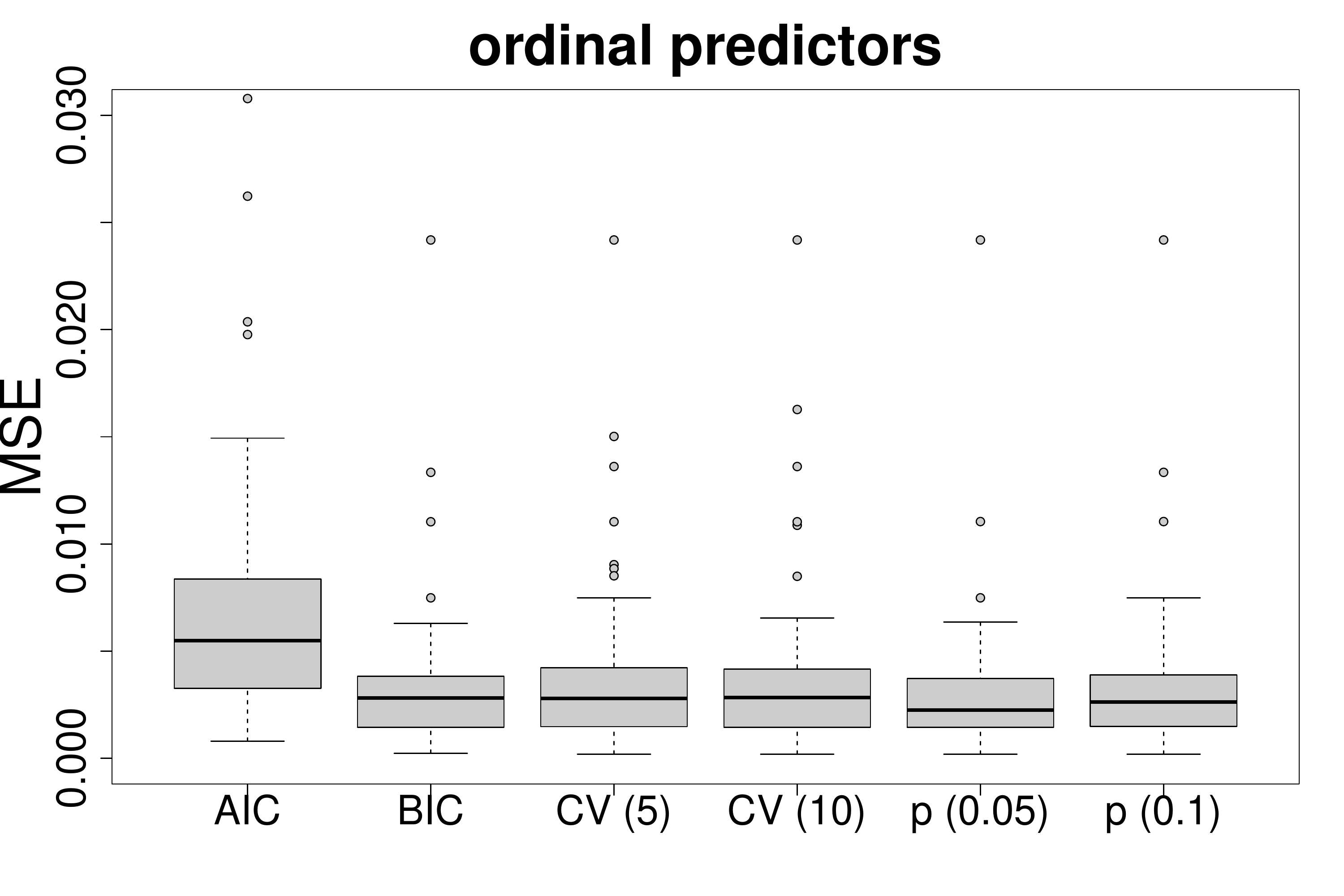}&\includegraphics[width=0.5\textwidth]{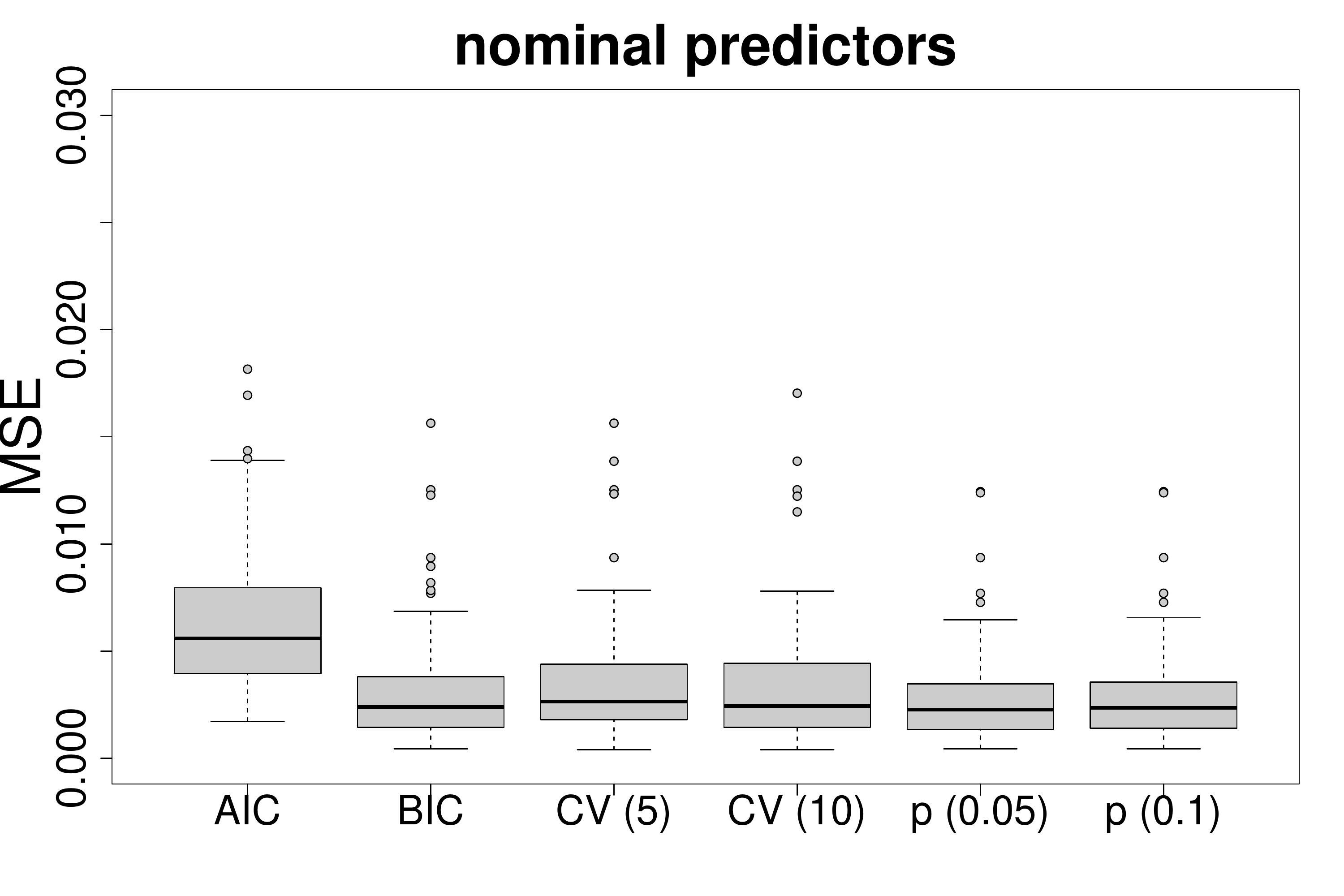}
\end{tabular}
\includegraphics[width=0.5\textwidth]{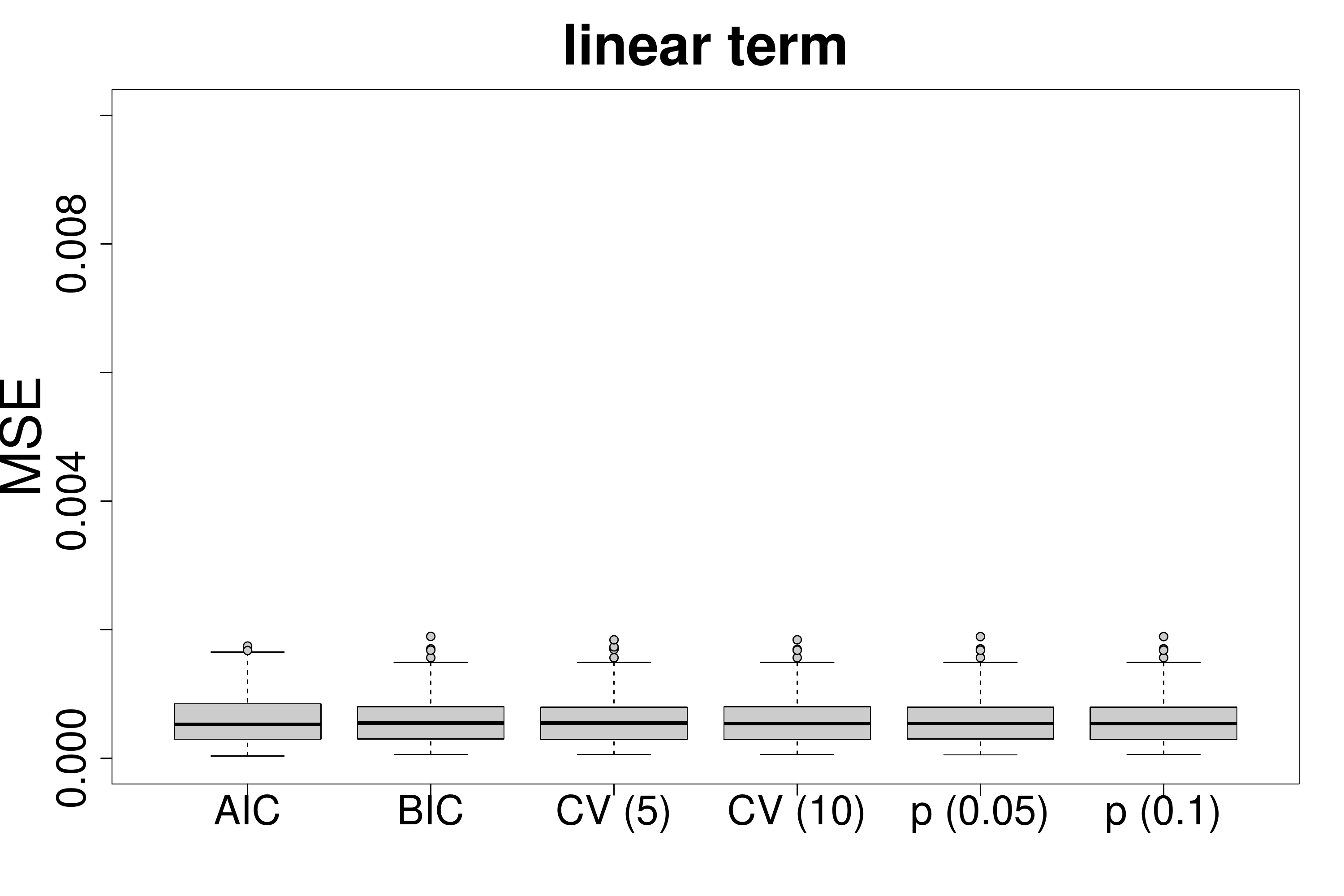}
\caption{Mean squared errors (MSEs) of parameter estimates of ordinal, nominal predictors and the linear term for the simulation study.}
\label{fig:BigSim_MSE}
\end{figure}

We consider the case of 4 ordinal and 4 nominal predictors in the tree component of our model. For both types of variables we use two predictors with 10 and two predictors with 5 categories. The true coefficients of the ordinal predictors are $(0,1,1,2,2,3,3,4,4)^\top$, $(0,0,0,0,2,2,2,2,2)^\top$, $(1,1,2,2)^\top$ and $(0,0,0,0)^\top$. For the nominal predictor they are $(0,0.5,0.5,-0.5,-0.5,1.5,1.5,-1.5,-1.5)^\top$, $(0,0,0,0,-2,-2,-2,-2,-2)^\top$, $(1,1,-1,-1)^\top$ and $(0,0,0,0)^\top$. In both cases the true numbers of clusters are 5, 2 and 3. The fourth predictor is not influential. Note that the effect of the first category in each case is set to zero. All in all there are 52 possible splits in the tree component. The true model contains 14 splits, 7 within the ordinal and 7 within the nominal predictors. We generate data sets with $n=2000$ observations and a normal distributed response with $\epsilon \sim N(0,1)$. Our model has an additional linear term $\xb^T\betab$, where $\xb$ is $N(\bf{0}_5,\Sigmab_5)$-distributed with variances 1 and covariances 0.3. The true regression coefficients of the linear term are $\betab=(-2,1,-1,3,2)^\top$.

The estimated coefficients are compared to the true parameters by calculating mean squared errors (MSEs). Therefore we distinguish between the tree-based parameters $\alphab$ and the parameters $\betab$ of the linear term. For  the $i$-th predictor  the MSE of the $\alpha$-parameters is $\sum_{j=1}^{m_i} (\hat{\alpha}_{ij}-\alpha_{ij})^2/m_i$ and for the $\beta$-parameters it is $\sum_{i=1}^{5} (\hat{\beta}_i-\beta_i)/5$.

In our analysis we distinguish between the MSE for the nominal and the MSE for the ordinal predictors respectively as the average over the four predictors. Boxplots of the computed MSEs based on 100 simulations are shown in Figure \ref{fig:BigSim_MSE}. We compare six different stopping criterions: AIC, BIC, 5-fold cross validation, 10-fold cross validation, $p$-values with significance level $\alpha=0.05$ and $p$-values with $\alpha=0.1$.  In Figure \ref{fig:BigSim_MSE} the latter are denoted by $p(0.05)$ and $p(0.1)$. The smallest median of MSEs for the ordinal predictors as well as for the nominal predictors were found for the strategy with p-values and common significance level $\alpha=0.05$ (fifth boxplots). MSEs for the linear term are very small and almost indentical over stopping criteria. Estimation of the linear term of the tree-structured model shows very good performance and seems to be not strongly linked to the clustering in the tree component.

\begin{figure}[!ht]
\centering
\begin{tabular}{cc}
\includegraphics[width=0.5\textwidth]{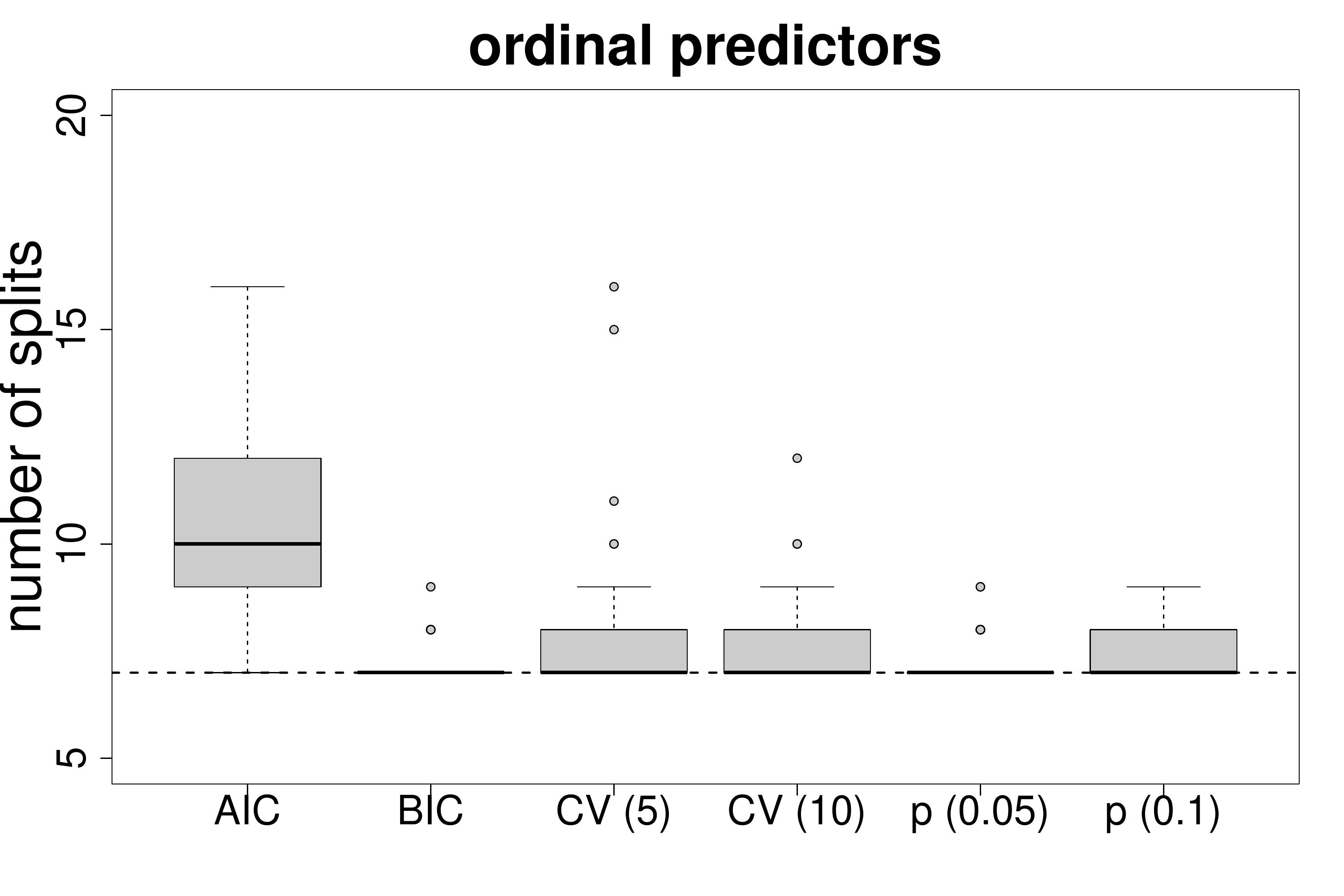}&\includegraphics[width=0.5\textwidth]{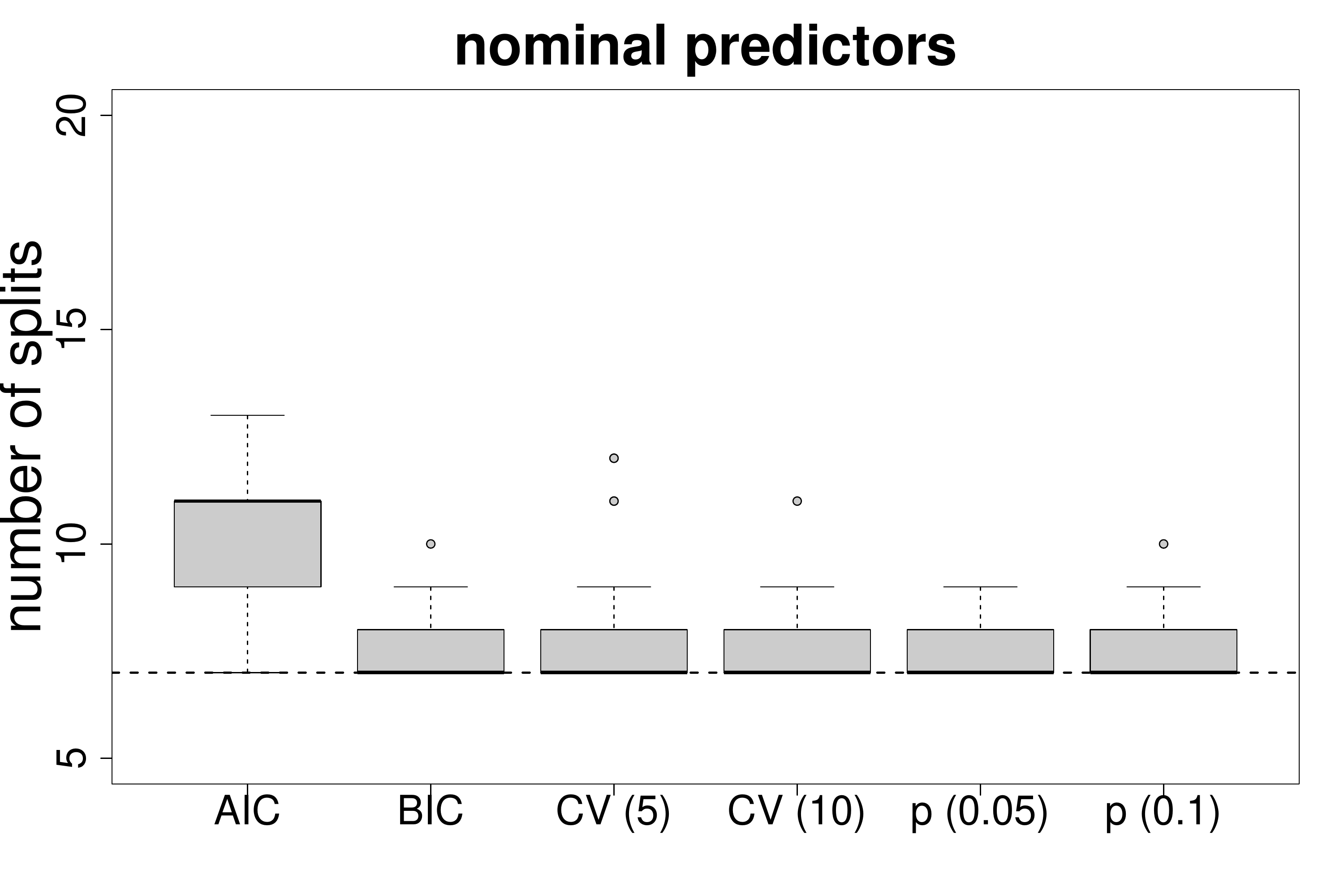}
\end{tabular}
\caption{Number of splits of ordinal and nominal predictors in the tree  component  for the simulation study.}
\label{fig:BigSim_ncluster}
\end{figure}

In addition to the MSEs the number of clusters found by the algorithm are of interest.
Therefore, Figure \ref{fig:BigSim_ncluster} shows the number of splits in the tree component of the model separately for the ordinal and the nominal predictors. The horizontal line shows the optimal number of splits of the underlying data generating model. It is seen that  for ordinal predictors one obtains nearly perfect results with BIC and $p$-value  $\alpha=0.05$. The true number of 7 splits is found in almost all simulations. For nominal predictors the  performance  is very similar over stopping criteria with the exception of AIC, which performs worse than the other procedures. Since  for $p$-values with $\alpha=0.05$ there is no outlier it shows again the best performance. In summary, the number of splits very close to the optimal number for all the procedures showing that the model is able to find the right number of splits.

\begin{figure}[!ht]
\centering
\begin{tabular}{cc}
\includegraphics[width=0.5\textwidth]{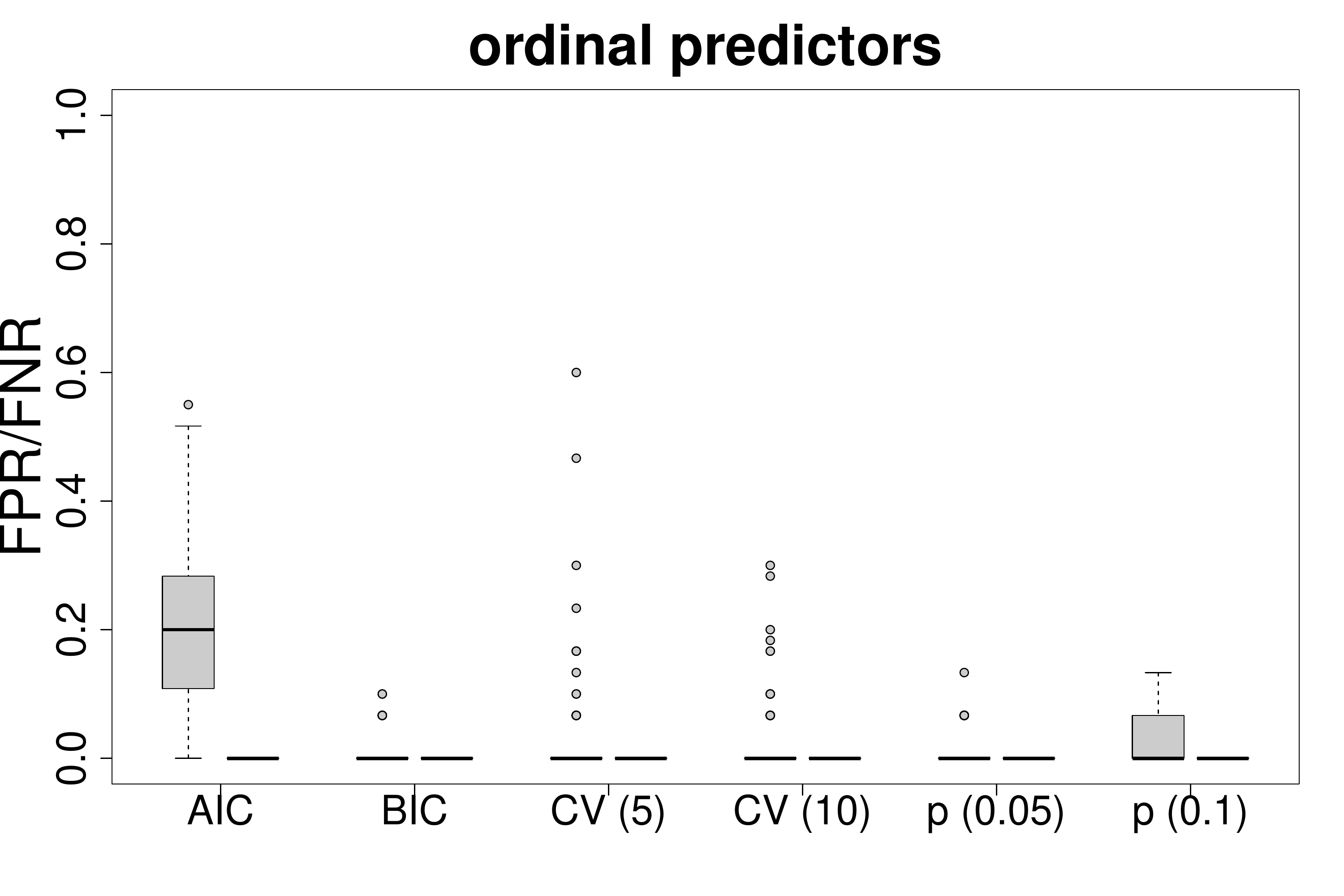}&\includegraphics[width=0.5\textwidth]{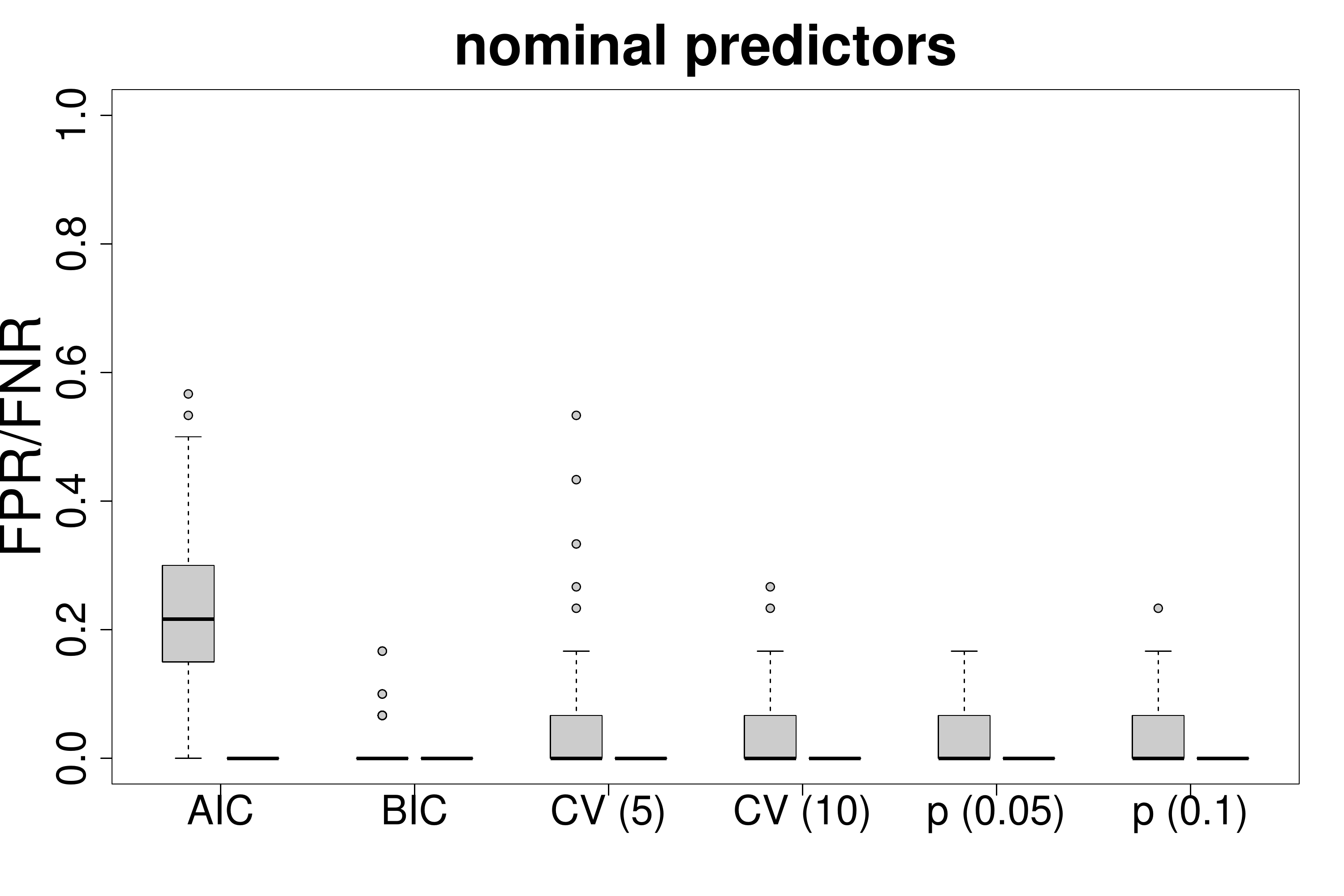}
\end{tabular}
\caption{FPR (left boxplots) and FNR (right boxplots) of ordinal and nominal predictors in the tree component  for the simulation study.}
\label{fig:BigSim_fpfn}
\end{figure}

To check if the algorithm does not only find the right number of splits but also the correct clusters, False Positive and False Negative Rates (FPR/FNR) are computed.

\begin{itemize}
\item False Positive: A difference between two estimated parameters $\hat{\alpha}_{ij}$ which is truly zero is set to nonzero
\item False Negative: A difference between two estimated parameters $\hat{\alpha}_{ij}$ which is truly nonzero is set to zero
\end{itemize}

Figure \ref{fig:BigSim_fpfn} shows boxplots of TPR and FPR seperatly for the ordinal and nominal predictors. As for MSEs we compute the average over the four predictors. Since the  tree-structured model has a weak tendency to overestimate the number of splits (see Figure \ref{fig:BigSim_ncluster}) FNRs are found to be zero in all simulations. With exception of AIC also the median of the FPRs is zero over stopping criteria.

In summary, the simulation study shows that the tree-structured model is able to identify the number of clusters rather correctly and yield very low false positive and negative rates. The algorithm  reduces the model complexity of categorical explanatory variables in the presence of smooth components very precisely.

\section{Further Applications}

\subsection{Car in household}

\begin{table}[!ht]
\centering
\begin{footnotesize}
\begin{tabularx}{\textwidth}{Xcrr}
\hline
\bf{Predictor}&\bf{Cluster}&\bf{Coefficient}&\bf{Stability}\\
\hline
&&&\\
Country&BE,HH,HB&-0.938&0.705\\
&NW,ST&-0.389&0.633\\
&BY,SH,NI,HE,RP-SL,BW,&&\\
&MV,BR,TH,SN&0.000&0.587\\
&&&\\
Number of persons&1&0.000&1.000\\
&2,3,4,5,6,7,8,9,11&1.493&0.573\\
&&&\\
Kind of household&8,3,2,5&-0.607&0.685\\
&6,4,7,1&0.000&0.767\\
\hline
\end{tabularx}

\vspace{0.5cm}

\begin{tabularx}{\textwidth}{Xrr}
\hline
\bf{Predictor}&\bf{Coefficient}&\bf{95$\%$ confidence interval}\\
\hline
&&\\
net income of all persons&0.0015&[0.0013,0.0016]\\
PC in household&1.0080&[0.6873,1.3724]\\
life policy during the year before&0.6910&[0.5715,0.8738]\\
\hline
\end{tabularx}
\end{footnotesize}
\caption{Estimated coefficients, stability measures of the tree component and $95\%$ confidence intervals of the linear term for the analysis of the Automobile data with a optimal number of four splits in the tree component.}
\label{tab:estimate_pkw}
\end{table}

As second application we consider data from the German socio-economic panel, which comprises 6071 households. The  response variable we consider is the binary variable if a car is in the household or not. Independent variables that we include in our model are the net income of all persons in the household in DM (metric),  the country (16 categories), type of household (nominal factor), number of persons in the household (ordinal factor), PC in household (yes/no), life insurance during the year before (available/not available).

\begin{wraptable}{r}{6cm}
\begin{scriptsize}
\begin{tabular}{r|l}
BW&Baden-Wuerttemberg\\
BY&Bavaria\\
BE&Berlin\\
BB&Brandenburg\\
HB&Bremen\\
HH&Hamburg\\
HE&Hesse\\
NI&Lower Saxony\\
MV&Mecklenburg-Vorpommern\\
NW&North Rhine-Westphalia\\
RP&Rhineland-Palatinate\\
SL&Saarland\\
SN&Saxony\\
ST&Saxony-Anhalt\\
SH&Schleswig-Holstein\\
TH&Thuringia
\vspace{-0.5cm}
\end{tabular}
\end{scriptsize}
\caption{German country code listed as in the ISO 3166-2.}
\label{tab:CC_DE}
\vspace{-0.5cm}
\end{wraptable}

A particularly interesting variable is the country. In the dataset the variable has 15 categories because Rheinland-Pfalz and Saarland are merged to one category. In a parametric model it generates 14 parameters. With the approach suggested here the number should reduce because it aims at identifying clusters of countries that share the same effect.

We fit a logistic regression model for the probability of holding a car and use $p$-values as
the stopping criterion. The tree component of the model includes the nominal factors country, type of household and the ordinal factor number of persons. The metric variable net income  and the two binary variables are put in the linear term of the model. The maximum number of splits in this case is 30. The algorithm stops very early and we obtain the model with four splits as the best model.

The results of the fitted tree-structured model are given in Table \ref{tab:estimate_pkw}, where the countries are abbreviated by the official country codes by ISO 3166 standard given in Table \ref{tab:CC_DE}. Table \ref{tab:estimate_pkw} shows in particular estimated coefficients, stability measures for clusters in the tree component and $95\%$ confidence intervals for the linear term based on 1000 bootstrap samples. It is seen that the three variables in the linear term all have a significant influence on the probability of having a car in the the household. The higher the net income, the higher the probability of holding a car. Also a PC in the household and  a life insurance  increase the probability of holding a car.

\begin{figure}[!ht]
\centering
\includegraphics[width=0.7\textwidth]{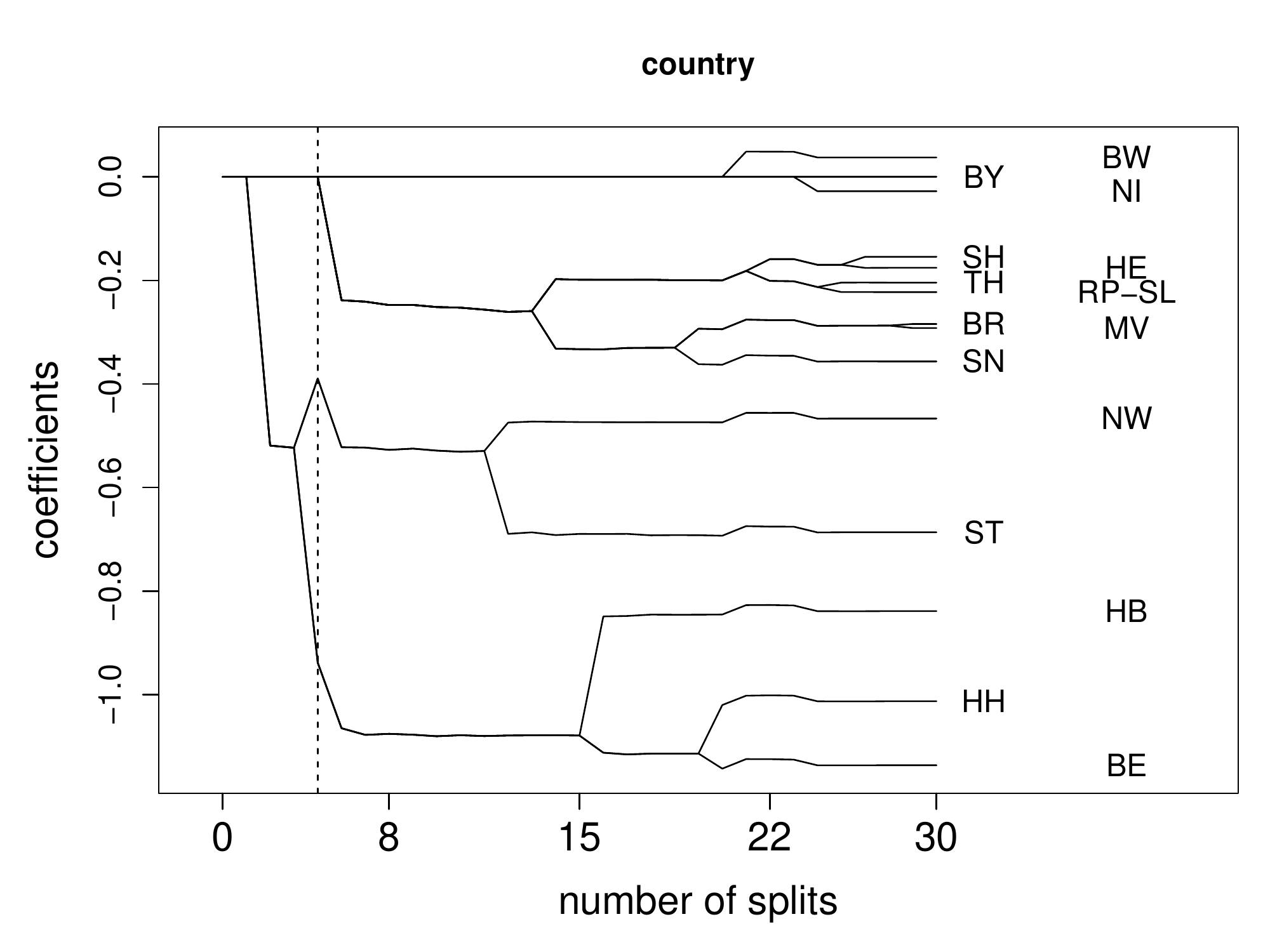}
\caption{Coefficient paths for the nominal predictor country for the analysis of the household data.}
\label{fig:bl_pkw}
\end{figure}

\begin{figure}[!ht]
\centering
\includegraphics[width=0.8\textwidth]{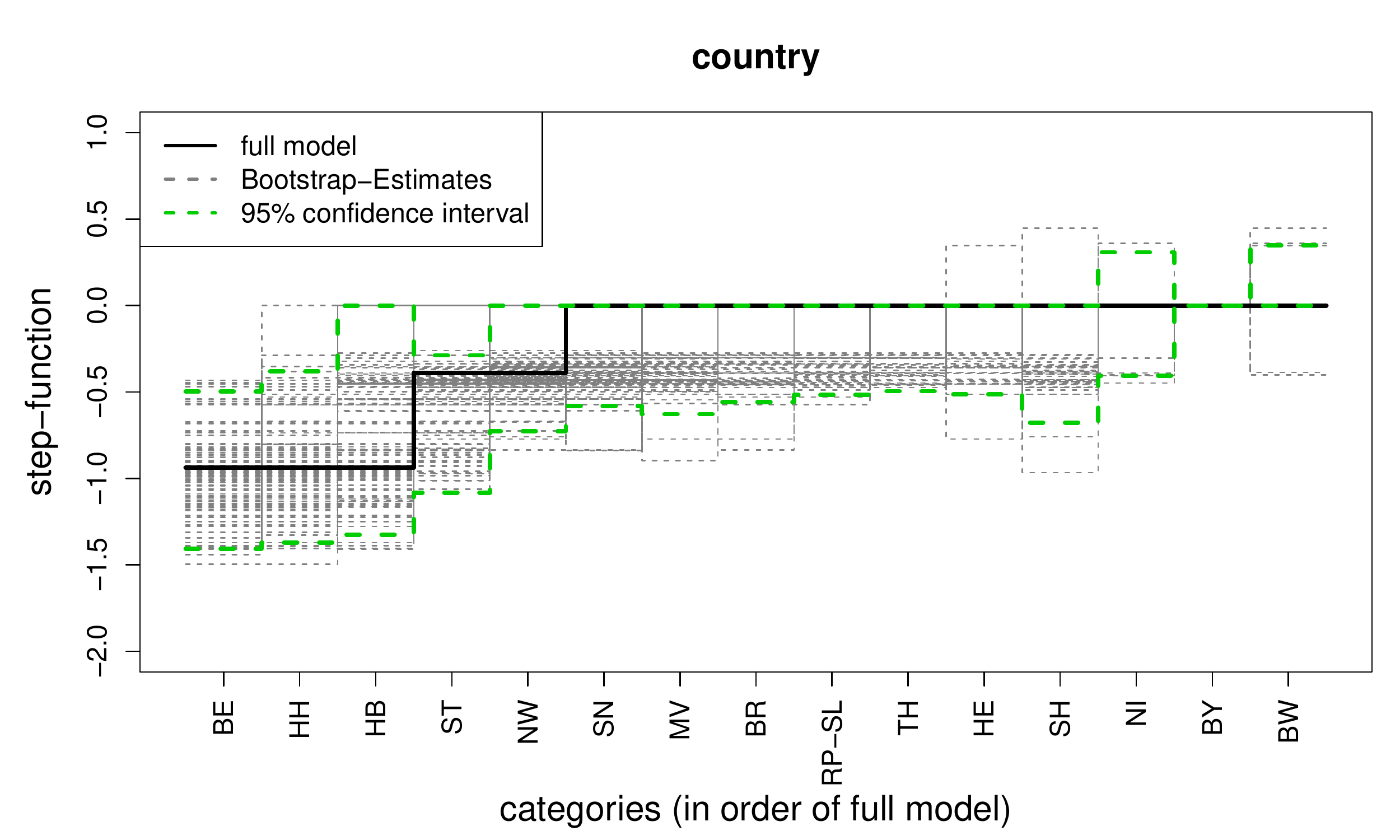}
\includegraphics[width=0.8\textwidth]{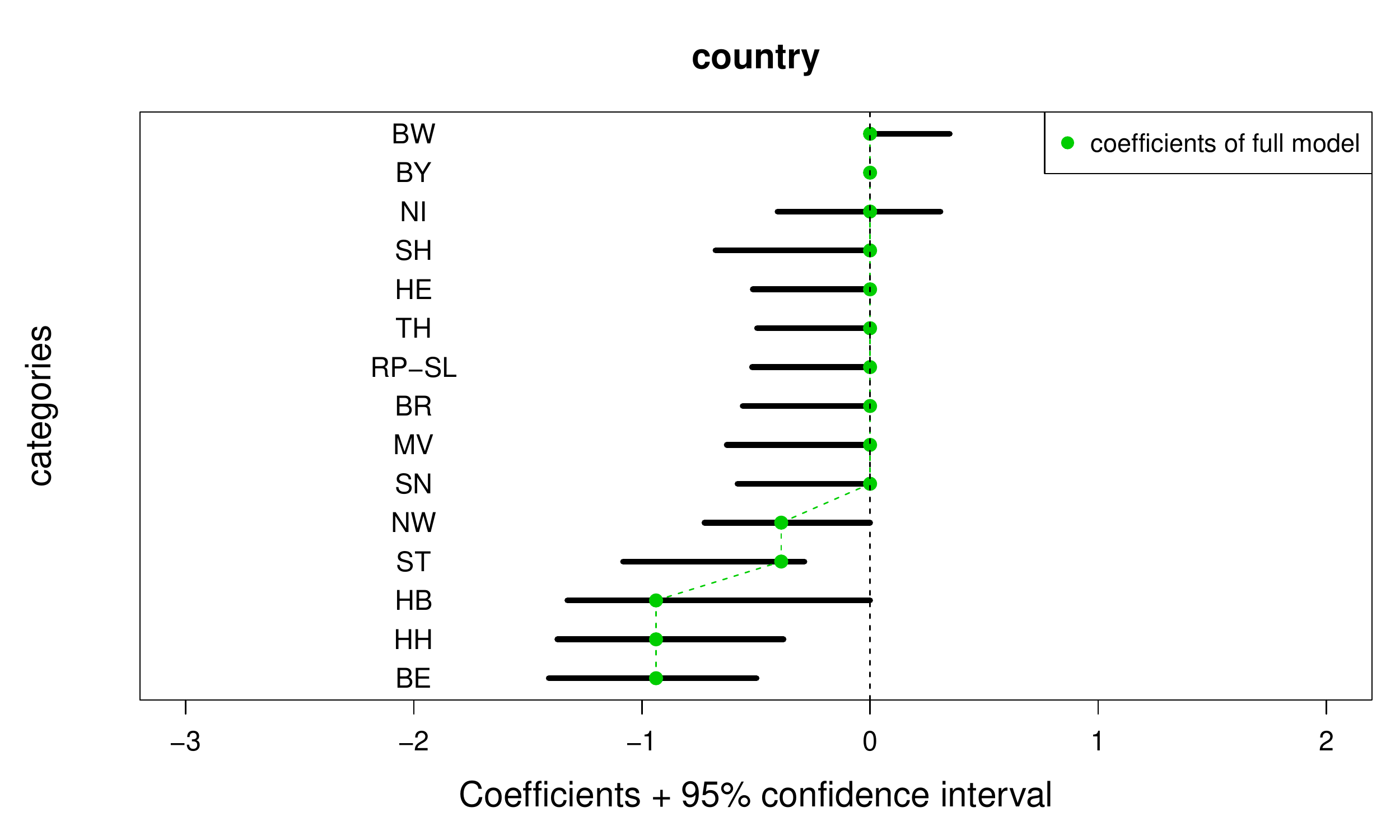}
\caption{Estimated step functions and resulting $95\%$ confidence intervals for the nominal predictor country for the analysis of the household data based on 1000 bootstrap samples.}
\label{fig:bln_pkw_KI}
\end{figure}

For  the nominal predictor country  in the tree component one obtains only three clusters that show an interesting structure. The first cluster, which has the strongest decrease in probability, is formed by the cities Berlin, Hamburg and Bremen, which are not only cities but also countries. Since in German cities public transportation is easily available and distances are small the necessity of owning a car given fixed income is reduced.  The coefficient  $-0.938$ means that the probability of owning a car decreases by a factor of $0.4$ when compared to the reference cluster with effect zero.  The smallest cluster contains only North Rhine-Westphalia and Saxony-Anhalt, which also have a reduced probability as compared to the rest, but the reduction is not as strong as for the countries that are also cities. As seen from the coefficient paths in Figure \ref{fig:bl_pkw} the big cluster could also divided into two sub-clusters, but were merged by the chosen stopping criterion.
For the number of persons in the household one obtains only two clusters. It is only distinguished between one person households and the rest of the households; households with more than one person show a strongly increased probability of owning a car. Stability measures in Table \ref{tab:estimate_pkw} for the predictor country are greater than 0.5 and do not vary a lot, so the algorithm forms stable clusters.

Figure \ref{fig:bln_pkw_KI} shows the fitted functions for 100 bootstrap samples and $95\%$ confidence intervals based on 1000 bootstrap samples for the predictor country . It is seen that the chosen reference Bavaria is not the first country in the  order of countries and so does not have the highest probability of outcome in the data. Only the confidence intervals of the two big states Baden-Wuerttemberg and Lower Saxony contain values greater than zero. The effects of the cities Hamburg and Berlin and Saxony-Anhalt are significantly different from zero. The bootstrap interval of Bremen is very large due to a small number of observations.

\subsection{Rating Scales}

The third application  concerns a comprehensive mood questionnaire, the so-called Motivational States Questionnaire (MSQ). It was developed to study emotions in laboratory and field settings. The data was collected between 1989 and 1998 at the Personality, Motivation, and Cognition Laboratory, Northwestern University (see \citet{rafaeli2006msq}). The data is part of the R package \texttt{psych} (\citet{revelle2013psych}). The original version of the MSQ included 70 items. Due to a huge number of missing values we use a revised version of 68 items of 1292 participants for our analysis. The response format was a four-point scale that asks the respondents to indicate their current standing with the following scale: 0 (not at all), 1 (a little), 2 (moderatly), 3 (very much).

\begin{figure}[!ht]
\centering
\includegraphics[width=1\textwidth]{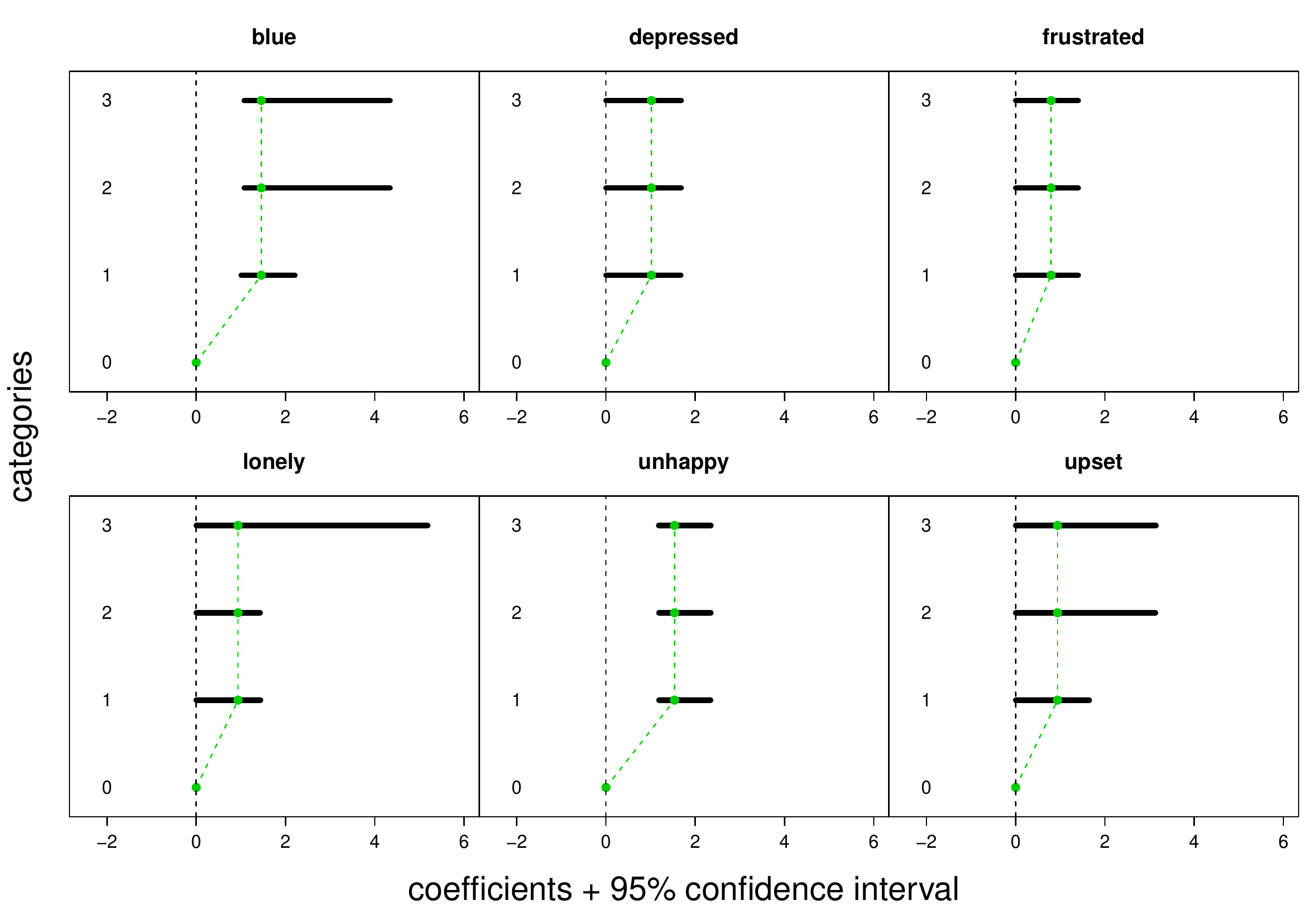}
\caption{Fitted coefficients of the full model (green dashed lines) and estimated $95\%$ confidence intervals based on 1000 bootstrap samples for the six items of the MSQ data that are included in the model.}
\label{fig:KI_rs}
\end{figure}

As response variable $y$ we consider the indicator if the participant feels sad or not, generated from the answers given for the item that asks for being \glqq said\grqq{}. The probability of feeling sad is modeled by a logistic regression model as in the hosehold data. The linear predictor consists of 67 ordinal predictors. Each predictor has four categories and corresponds to one item that was asked for in the questionnaire. There is no additional linear term in our model. The example illustrates that the approach is able to handle a large number of ordinal predictors.

The fitted coefficients and estimated $95\%$ confidence intervals based on 1000 bootstrap samples for the predictors that are included in the model are shown in Figure \ref{fig:KI_rs}. It  is seen that only six variables among the 67 available variables were selected. Only the items that ask for being \glqq blue\grqq{}, \glqq depressed\grqq{},\glqq  frustrated\grqq{}, \glqq lonely\grqq{}, \glqq unhappy\grqq{} and \glqq upset\grqq{} are considered as being influential. Moreover, there is substantial clustering of the categories of the predictors. The coefficients of each predictor is a constant for level 1 to 3 reducing the ordinal predictors to binary predictors that distinguish between category 0 and the rest only.  Bootstrap based confidence intervals are not the same for levels 1 to 3 in each case. Hence, there are bootstrap samples where the clusters consisting of level 1 to 3 are split a second time. Only for emotions \glqq blue\grqq{} and \glqq unhappy\grqq{} the confidence intervals do not contain zero.
Thus it can be concluded that there are only 2 out of 67 emotions that have a significant effect on the probability of being sad.

\subsection{Comparison with Alternative Models}

\begin{figure}[!ht]
\centering
\includegraphics[width=0.5\textwidth]{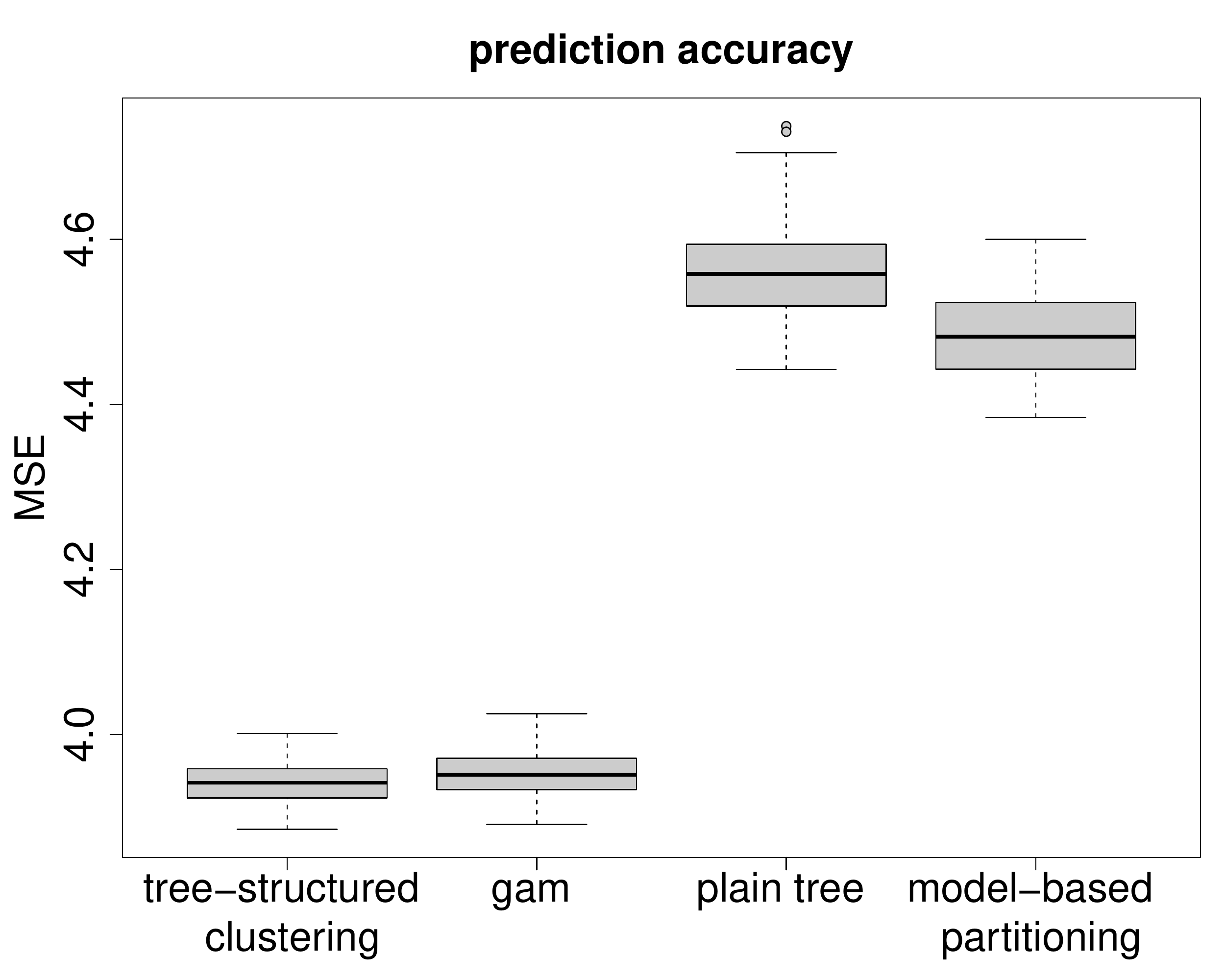}
\caption{Comparison of prediction accuracy of tree-structured clustering with other methods for the Munich rent data.}
\label{fig:miete_vgl}
\end{figure}

\begin{figure}[!ht]
\centering
\includegraphics[width=0.5\textwidth]{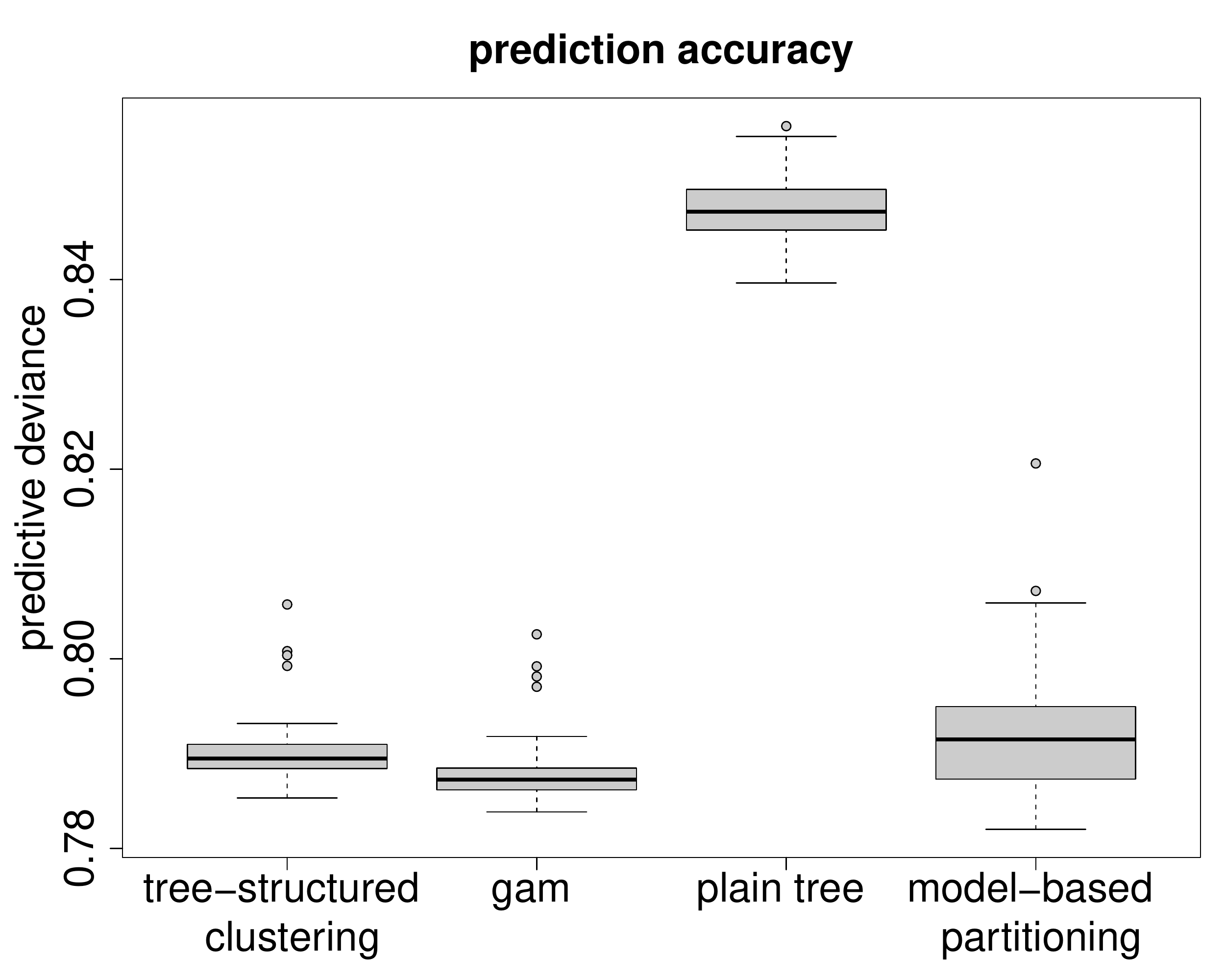}
\caption{Comparison of prediction accuracy of tree-structured clustering with other methods for the household data.}
\label{fig:pkw_vgl}
\end{figure}

In the previous sections the tree-based model was used to identify clusters in categorical predictors. Although prediction is not the main objective of the modeling strategy  one expects  any appropriate model to also perform well in terms of prediction accuracy. Therefore, we briefly compare  the tree-based model with its main competitors with regard to prediction accuracy.
Since in simulations typically one model, namely the data generating model, is preferred  we consider the performance for the real data sets.  The predictive deviance in both cases was measured by 5-fold cross-validation using 100 repetitions. As competing models we used the generalized additive model, a plain tree and model based partitioning. The generalized additive model  was estimated by function gam from package mgcv \citep{wood2011mgcv}.
The plain tree was estimated by use of the function rpart from package rpart \citep{therneau2014rpart}. The complexity parameter 'cp' determines the minimal reduction of lack of fit. The optimal parameter was found to be 0.01 in both examples. Model based partitioning  was estimated by the function mob of package party \citep{zeileis2008model}. Predictors in the tree component of our model were used for partitioning. Predictors in the parametric part of our model were passed to models in each leaf. Complexity parameter 'trim', specifies the trimming in the parameter instability test. The optimal parameter was found to be 0.05 (rent) and 0.03 (car).
Figures \ref{fig:miete_vgl} and \ref{fig:pkw_vgl} show the results for the rent data and the household data, respectively. It is seen that the tree-based model and GAM have comparable performance, which was to be expected since the tree-based model is essentially a GAM but with built-in clustering. The plain tree, with its focus on interaction shows much worse performance whereas model based partitioning performs poorly in one case and rather well in the other case.

\section{Concluding Remarks}

The proposed tree-structured approach is a modelling tool that allows to identify clusters in categorical predictors for nominal and ordinal predictors. In particular when several predictors with potentially many categories are available it is an efficient tool to reduce the superfluous complexity of classical parametric models. Simulation results show that the algorithm works well.

It should be noted that the tree-structured approach does not yield a tree in the sense of traditional recursive partitioning, where models are fitted recursively to sub samples defined by nodes. In the tree-structured model one obtains for each of the categorical predictors that are used in the tree component a separate tree. The obtained trees show which categories have to be distinguished given the other predictors are included in the model.

\bibliography{literatur}

\end{document}